\documentclass{aa}  
\usepackage{graphicx}
\usepackage{multirow}
\usepackage{dcolumn}
\usepackage{txfonts}

\usepackage{natbib}
\usepackage{mathrsfs}
\usepackage{lineno}
\linenumbers

\usepackage[draft=false]{hyperref}
\hypersetup{colorlinks,linkcolor={red},citecolor={blue},urlcolor={blue}}

\newcommand{\dd}{{\rm d}}
\newcommand{\msun}{$M_{\sun}$}

\begin{document}
\title{Reconstruction of spider system's observables from  orbital period modulations via the Applegate mechanism}
\titlerunning{Reconstruction of spider system's observables}
\authorrunning{De Falco et al. (2024)}
\author{Vittorio De Falco\inst{1,2}\thanks{\email{v.defalco@ssmeridionale.it}}, Amodio Carleo\inst{3}\thanks{\email{amodio.carleo@inaf.it}},
Alessandro Ridolfi\inst{3} \thanks{\email{alessandro.ridolfi@inaf.it}}, Alessandro Corongiu\inst{3}\thanks{\email{alessandro.corongiu@inaf.it}}
}
\institute{Scuola Superiore Meridionale, Largo San Marcellino 10, 80138 Napoli, Italy,
\and
Istituto Nazionale di Fisica Nucleare (INFN), sez. di Napoli, Via Cinthia 9, I-80126 Napoli, Italy
\and
INAF – Osservatorio Astronomico di Cagliari, Via della Scienza 5, I-09047 Selargius (CA), Italy
}

\date{Received \today; accepted XXX}

\abstract{Redback and black widow pulsars are two classes of peculiar binary systems characterized by very short orbital periods, very low mass companions, and, in several cases, regular eclipses in their pulsed radio signal. Long-term timing revealed systematic but unpredictable variations in the orbital period, which can most likely be explained by the so-called Applegate mechanism. This relies on the magnetic dynamo activity generated inside the companion star and triggered by the pulsar wind, which induces a modification of the star's oblateness (or quadrupole variation). This, in turn, couples with the orbit by gravity, causing a consequent change in the orbital period. The Applegate description limits to provide estimates of physical quantities by highlighting their orders of magnitude. Therefore, we derive the time-evolution differential equations underlying the Applegate model, that is, we track such physical quantities in terms of time. Our strategy is to employ the orbital period modulations, measured by fitting the observational data, and implementing a highly accurate approximation scheme to finally reconstruct the dynamics of the spider system under study and the relative observables. Among the latter is the magnetic field activity inside the companion star, which is still a matter of debate for its complex theoretical modeling and the ensuing expensive numerical simulations. As an application, we exploit our methodology to examine two spider sources: 47 Tuc W (redback) and 47 Tuc O (black widow). The results obtained are analyzed and then discussed with the literature. }

\keywords{Stars: binaries: eclipsing, Stars: magnetic field, Binary Systems}

\maketitle

\section{Introduction}
Among the more than $\sim3600$ radio pulsars currently known\footnote{See \url{https://www.atnf.csiro.au/research/pulsar/psrcat}, for more details. However, the situation is as follows: within the galactic plane, 135 are isolated and 209 are binaries; whereas in globular clusters, 87 are isolated and 91 are binaries.}, about $14\%$ are millisecond pulsars (MSPs). These are old neutron stars (NSs) endowed with relatively weak magnetic fields ($B\simeq10^7-10^9$ G) and very short rotational periods ($\sim1-10$~ms). MSPs represent the progeny of low-mass X-ray binaries (LMXBs), where the NS gets spun up by the accreted matter (coming from the secondary star via Roche lobe overflow) and finally attains extreme rotation rates. This is known and widely accepted in the literature as the \emph{recycling scenario} \citep{Bisnovatyi1974,Alpar1982,Radhakrishnan1982,Bhattacharya1991,Papitto2013}.

Although MSPs are formed in binary systems, about $20\%$ of them are isolated \citep{Belczynski2010}. The reason for this is still unclear and is matter of debate \citep{Heuvel1988,Rasio1989,Bhattacharya1991}. A possible explanation arose after the discovery of the first eclipsing binary pulsar, PSR B1957+20 \citep{Fruchter1988}, in which the companion star is constantly ablated by energetic particles and/or $\gamma$-rays produced by the pulsar wind \citep{Kluzniak1988,Heuvel1988,Ruderman1989}. This led astronomers to propose the so-called \emph{evaporation scenario}, accordong to which the secondary star gets ablated until it fully disappears, thus leaving an isolated MSP. However, successive estimates of the mass-loss rate showed that the evaporation time scale is likely much longer than the Hubble time, casting doubts on the effective occurrence of this phenomenon \citep{Stappers1996a,Stappers1996b,Stappers2001}.

Notwithstanding, eclipsing binary pulsars became increasingly important for stellar and binary evolution studies. The recycling model, initially supported by the observation of accreting millisecond X-ray pulsars \citep[AMXPs; see e.g.,][]{Wijnands1998,Falanga2005}, was ultimately confirmed by the discovery of three ``transitional''  pulsars, (PSR J1023$-$0038, J1824$-$2452I, and J1227$-$4853), i.e. systems that have been observed swinging between radio-MSP and X-ray binary states \citep[e.g.][]{Papitto2013,Stappers2014}.

\emph{Spider pulsars} are a subclass of binary MSPs, characterized by tight (orbital period $\lesssim1$ d) and circular (eccentricities $\simeq10^{-3}-10^{-4}$) orbits \citep[see e.g.,][] {Romani2012,Pallanca2012,Kaplan2013}, and a light-weight companion. Most (but not all) of them are also eclipsing binary pulsars.
Depending on the mass $m_c$ of the companion, they can be further divided into two distinct classes \citep{Roberts2013}: \emph{black widows} have degenerate companions with  $m_c\lesssim0.1M_\odot$, whereas \emph{redbacks} have semi-degenerate companions with $m_c\simeq0.1-0.4M_\odot$. The evolutionary scenario of these two types of binary systems has long been discussed, and it is now widely accepted to occur through \emph{irradiation processes} \citep[][and references therein]{Podsiadlowski1991,DAntona1993,Bogovalov2008,Bogovalov2012,Chen2013,Smedley2015}.

The long-term timing of several spiders has revealed that they often show significant modulations of their orbital periods, and sometimes also of their projected semi-major axis \citep[see e.g.,][]{Shaifullah2016,Ng2018}. These variations generally manifest themselves as recurrent, but not strictly periodic, cycles. These observational clues lead to the exclusion, as dynamical explanations, of apsidal motions \citep{Sterne1939} and presence of third bodies \citep{Buren1986}. Instead, a plausible reason can be the \emph{magnetic activity inside the companion star} \citep{Hall1990}, closely linked to the dynamo action due to the presence of differential rotation and convective zones. 

Spider systems are usually found in \emph{quasi-tidally-locked configurations}, occurring when there is no relevant angular momentum transfer between the companion star and its orbit around the pulsar. This is due to the tidal force acting between the co-orbiting bodies through the pulsar irradiation-driven winds \citep[see e.g.,][]{Applegate1994,Bogdanov2005}, entailing tidal dissipation of the pulsar on the companion \citep[known as \emph{tidally-powered star}; see e.g.,][]{Balbus1976,Kochanek1992,Zahn2008} and \emph{synchronous rotation} (one hemisphere of a revolving body constantly faces its partner). 

A possible explanation for the orbital period variations relies on the \emph{Applegate mechanism} \citep{Applegate1987,Applegate1992pro,Applegate1992}, where magnetic cycles induce deformations on the companion star shape, thus altering its quadrupole moment, consequentially causing gravitational acceleration and orbital period modulations. This phenomenon is triggered by the irradiation-driven winds from the pulsar, which generates a spin torque on the companion. This in turn induces tidal dissipation and energy flow, which powers the magnetic dynamo \citep{Applegate1994}. 

The observed orbital period and projected semi-major axis modulations can be influenced by other effects that are intrinsic to the system or caused by kinematic reactions relative to the observer motion (e.g., emission of gravitational waves, Doppler corrections, mass-loss of the binary, and tidal bulge forces), which \citet{Lazaridis2011} estimated to be orders of magnitudes smaller than those caused by the gravitational quadrupole moment activity. On the other hand, \citet{Lanza1998,Lanza1999} proposed an alternative explanation to the gravitational quadrupole moment variations. Their model applies the \emph{tensor virial theorem} \citep{Chandrasekhar1961} to a general magnetic field geometry to formalize, through an integral approach, the variations in oblateness. This implies a distributed non-linear dynamo in the convective envelopes of the companion star, which affects not only the quadrupole moment, but also the differential rotation. 

The Applegate mechanism is still the most quoted explanation for its good agreement with the observations, whose measured amplitudes of period modulations are $\Delta P/P\sim10^{-5}$ (over timescales of decades or longer), companion star's variable luminosity over the $\Delta L/L\sim0.1$ level, and differential rotations at the $\Delta \Omega/\Omega\sim0.01$ order. These values are common in spider systems \citep[see, e.g.][]{Ridolfi2016,Freire2017}.

More recently, \citet{Voisin2020a,Voisin2020b} improved the Applegate picture, also taking into account relativistic corrections. They proposed a detailed model for describing the motion of spider binary systems that allows us to accurately estimate the observed $\Delta P$ variations with the final objective of improving the timing solution of these gravitational sources.

The necessity in having solid theoretical assessments to describe spider systems represents a powerful means to: $(i)$ better understand the stellar magnetic activity, $(ii)$ get insights into the dynamo processes, $(iii)$ obtain more information on the companion stars' equation of state. In particular, the structure and generation of the magnetic field in low-mass stars are still not clear and need to be investigated. The \emph{surface} magnetic field is thought to be of the order of several kG and can be directly observed \citep[see Fig. 1 in][]{Han_2023}. Instead, the \emph{interior} fields are based on contrasting theoretical analyses strongly depending on the considered model \citep[see e.g.,][]{Rakesh,Gregory,macdonald}.

In this work, we employ the Applegate mechanism, which besides describing the phenomenology behind the spider pulsar, it also provides an estimate of the order of magnitude of some related physical variables (e.g., luminosity, differential rotations, quadrupole moment etc.). We propose a \emph{dynamical formulation of the Applegate model}, where the gravitational source and physical quantities' dynamics are tracked point by point during their time evolution. In addition, we exploit an \emph{inverse approach}, being counter-current with respect to the strategies followed in the latest papers \citep[see e.g.,][]{Voisin2020a,Voisin2020b}. Indeed, rather than finding a physical justification for the $\Delta P$ variations, we take its profile from the long-term observations to reconstruct the gravitational source and related observables' dynamics. The approach we adopt in this paper follows the \texttt{BTX} phenomenological model \citep[extension of Blandford \& Teukolsky (\texttt{BT}) model,][]{Blandford1976,Bochenek2015}, which allows one to fit the observational data within a precise timing baseline. This scheme gives rise to a set of coupled ordinary differential equations with respect to time, involving the orbital separation and quadrupole moment. However, the ensuing dynamical system is still difficult to solve analytically. Therefore, we also develop a mathematical procedure that provides highly accurate approximate analytical solutions. This methodology allows for an easy accomplishment of the proposed goals.

The paper is structured as follows: in Sec. \ref{sec:model}, we describe the features of our dynamical model and derive the equations of motion; in Sec. \ref{sec:approx_meth}, we propose a reasonable approximation pattern to infer an analytical solution; in Sec. \ref{sec:results} our achievements are applied to black widow and redback systems; finally, in Sec. \ref{sec:end} we conclude with some discussions and future perspectives.

\section{The model}
\label{sec:model}
We present the dynamical version of the Applegate mechanism, which reports the order of magnitude of the physical observables underlying the dynamics of black widow and redback binary systems. This model is further enhanced by incorporating orbital period modulations' profile, derived from observations. This gives rise to a set of coupled and non-trivial ordinary differential equations. We also introduce the explicit formulae of some physical variables, whose plotted profiles are the goals of our work.    

\begin{figure*}[ht!]
    \centering
    \includegraphics[trim=0.2cm 3cm 0.2cm 3cm,scale=0.57]{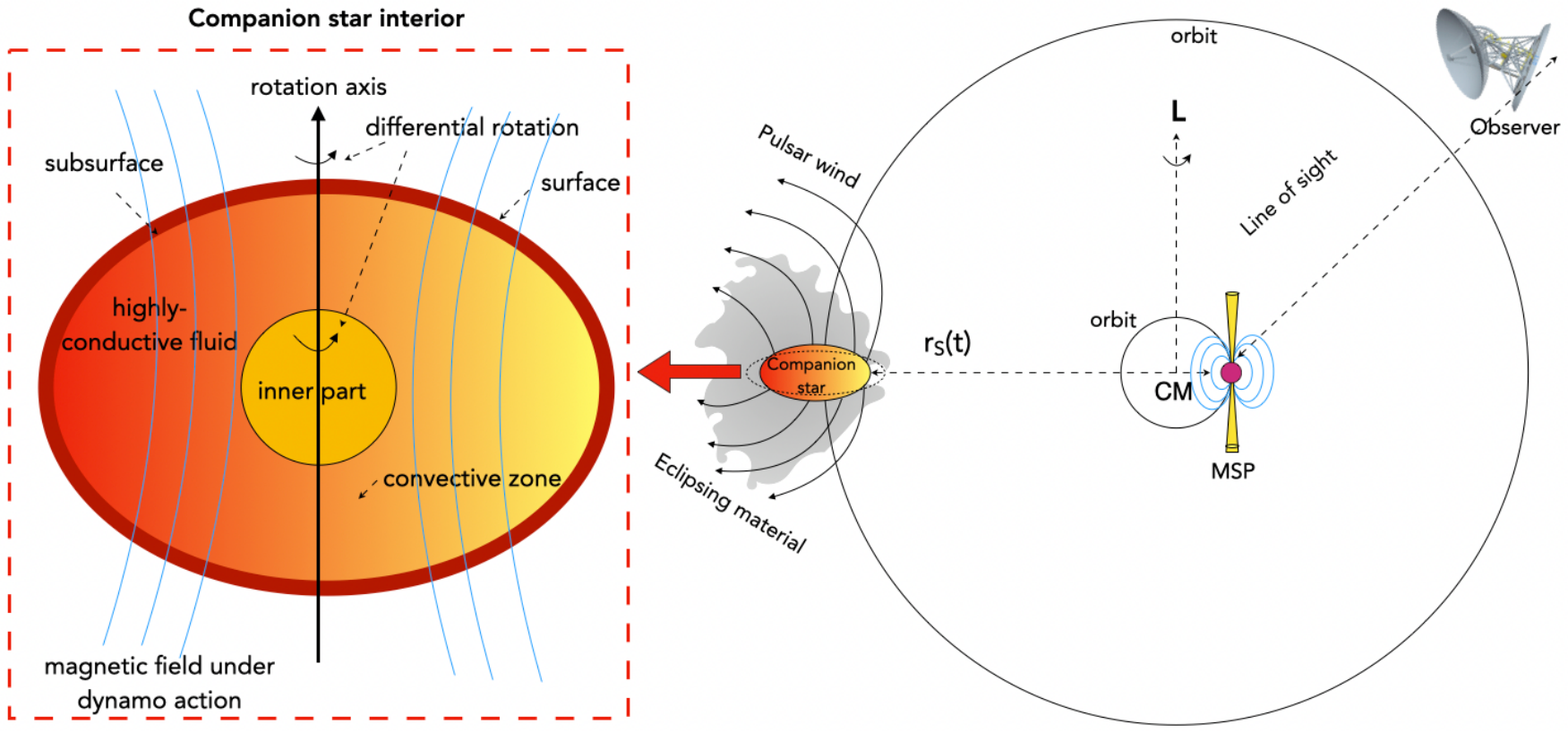}
    \caption{Illustration of a typical spider binary system. A MSP and a companion star are in a tidally locked configuration, moving in synchronous rotations on quasi-circular and tight orbits. They are separated by a distance $r_{\rm s}(t)$, in a plane orthogonal to the orbital angular momentum vector $\boldsymbol{L}$, which is conserved during the binary system's motion. The irradiation-driven wind from the MSP heats through a tidal dissipation the companion star, which starts to evaporate, loosing thus mass. This is the eclipsing material, which obstacles the radio signal detected by a telescope located far from the binary system. In the dashed red box we sketch the companion star interior. This zone is characterized  by a convective envelope, where the fluid in it is highly conductive. Furthermore, the pulsar wind triggers the differential rotation of the companion star, which induces a magnetic activity through a dynamo action. The subsurface magnetic field is responsible to break the hydrosthatic equilibrium inside the star, inducing quadrupole moment changes. Notice that the external magnetic field is expected to be locally poloidal, while the subsurface field is toroidal.}
    \label{fig:Fig1}
\end{figure*}

We deal with binary sources composed of a pulsar of mass $m_p$ and a companion star of mass $m_c$ and radius $R_c$. The two bodies are both treated as test particles, even if the companion star should be considered as extended to account for changes in shape. However, to simplify the mathematical treatment, we still regard this object as a test body and we entrust the quadrupole variable to characterize the variations of matter distribution within the star. In Fig. \ref{fig:Fig1} we report a cartoon sketching the geometry of the problem under investigation.

The definitions of some of the above parameters require additional clarifications. The companion star's mass, $m_c$, is not strictly constant over time, as it gradually decreases due to the mass loss driven by the pulsar wind. However, the fraction of matter lost over the observational period is so small (around $10^{-10} M_\odot/{\rm yr}$)\footnote{In this investigation, three aspects must be considered: (1) the mass loss from the companion amounting to $10^{-10} M_\odot/{\rm yr}$ \citep{Pan2023}; (2) if the Roche lobe is less than the radius of the companion star, there is mass transfer; (3) the pulsar wind entails a mass loss of $10^{-12} M_\odot/{\rm yr}$ \citep{Guerra2024}. We conclude that we can consider constant mass.} that it is reasonable to approximate $m_c$ through a constant value. Similarly, the radius of the companion star, $R_c$, is not fixed; it varies as the shape of the star changes due to quadrupole variations. Therefore, the definition of $R_c$ refers to the companion star's radius at rest. These fluctuations in the size dimension can be quantified, and their relative magnitude is estimated to be about $5-7\%$ with respect to $R_c$ \citep{Applegate1992}\footnote{The companion star's variation are due to the combination of two effects: quadrupole changes amounting to $1-5\%$ \citep{Applegate1992,Harvey1995} and magnetic field dynamo activity causing fluctuations of $2-6\%$ \citep{Rappaport1983,MacDonald2009}. The combined effects can thus range in the interval $5-7\%$.}.

This section opens by briefly recalling the Applegate mechanism in Sec. \ref{sec:Appmec}, representing the core of our work. In the construction of the model, we will make use of two reference frames (RFs) in Sec. \ref{sec:observers}, which allow us to conveniently deal with the ensuing dynamics. The equations of motion are presented in Sec. \ref{sec:EOMs}. We conclude by deriving the formulae of some fundamental physical observables involved in this scenario in Sec. \ref{sec:observables}. 

\subsection{Applegate mechanism and dynamo action}
\label{sec:Appmec}
\citet{Applegate1992} proposed a mechanism to explain the orbital period modulations in eclipsing binary systems as a consequence of the magnetic activity inside the secondary star, triggered by the pulsar wind. For spider binary systems (i.e., black widows and redbacks) the lighter body plays the role of the \emph{active star}. The main idea underlying this approach is based on the magnetic activity cycle, which represents the engine that causes a redistribution of the angular momentum inside the star, thus modifying  its oblateness  \citep[see e.g.,][]{Warner1988,Lanza2002,Donati2003,Lanza2006,Bours2016}. This induces a variation in the radial component of the gravitational acceleration via the \emph{gravitational quadrupole-orbit coupling}, thus entailing the orbital period modulations, which we eventually detect.

This magnetic activity seems to be powered by a \emph{dynamo action}, i.e., a process of magnetic field generation through the inductive response of a highly-conductive fluid. Indeed, there is a conversion of mechanical energy into a magnetic one by stretching and twisting the magnetic field lines \citep{Parker1955}. All of the aforementioned effects are driven by the subsurface magnetic field, located within the star's convective zones. 

The dynamo action is the alternation of two phenomena inside the star: (1) the sheared differential rotation at different latitudes contributes to the trasformation of an initially poloidal magnetic field into an enhanced toroidal one through the \emph{Alfvén theorem} \citep[also known as \emph{$\Omega$-effect}; see e.g.,][]{Parker1955,Parker1979,Browning2006}; (2) the combined action of cyclonic convection, buoyancy, and Coriolis forces turns the toroidal magnetic field back to the poloidal one, completing thus the cycle
\citep[also known as \emph{$\alpha$-effect}; see e.g.,][]{Parker1955,Parker1979,Choudhuri1995,Charbonneau2000,Browning2006}. 

The dynamo model is usually gauged on the solar magnetic activity, as it shares profound similarities with the problem under investigation. We know from Sun's observations that the period of the sunspot cycles (connected to the solar subsurface magnetic activity) is about 11 yr \citep{Baliunas1985}, but this value can be of much longer duration in spider systems. From this consideration and since various cycles can have different durations, we can conclude that \emph{the magnetic activity leads to systematic, but not strictly periodic, changes in the active star, subsequently causing the observed orbital period variations}.

\subsection{Reference frames}
\label{sec:observers}
Spider systems are fairly tight, implying highly-circular orbits. Furthermore, the tidal friction predominately acts, entailing a synchronization of spin and orbit of the binary system, i.e., rotational and orbital angular momentum vectors are aligned and also have the same module. The dynamics occurs in the plane $\mathcal{P}$ orthogonal to the direction of the orbital angular momentum. We neglect any additional perturbing effects that could potentially lead to three-dimensional motions outside the $\mathcal{P}$ plane.

\begin{figure}[ht!]
    \centering
    \includegraphics[trim=2.8cm 3cm 0cm 2.8cm,scale=0.35]{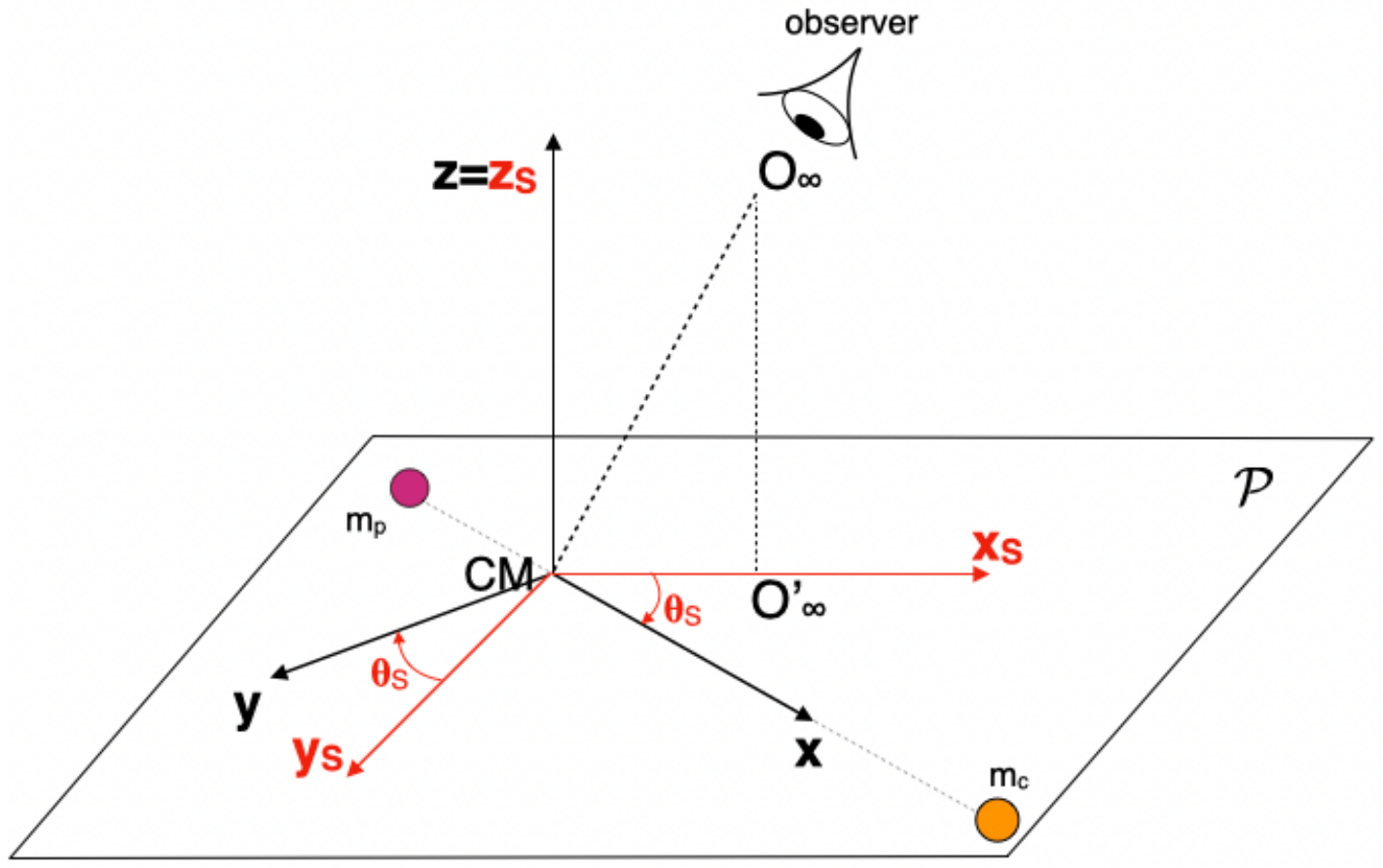}
    \caption{We display the two RFs: $\mathcal{R_{\rm CO}}$ (black) and $\mathcal{R_{\rm S}}$ (red).}
    \label{fig:Fig2}
\end{figure}

The dynamics can be described in two RFs, having both the origin in the binary system's center of mass (CM; see Fig. \ref{fig:Fig2}): 
\begin{itemize}
    \item \emph{orthonormal corotating RF}, $\mathcal{R}_{\rm CO}=\left\{\boldsymbol{x},\boldsymbol{y},\boldsymbol{z}\right\}$: the $\boldsymbol{z}$-axis is orthogonal to $\mathcal{P}$, where the $\boldsymbol{x}$- and $\boldsymbol{y}$-axes are placed. The $\boldsymbol{x}$-axis is always directed along the line connecting the two bodies and pointing towards the companion star. This RF co-rotates with the binary system, so the motion in it results to be always static. The only variations occur along the $\boldsymbol{x}$-axis, reducing thus the whole dynamics to just one dimension;
    
    \item \emph{orthonormal static RF}, $\mathcal{R}_{\rm S}=\left\{\boldsymbol{x_{\rm S}},\boldsymbol{y_{\rm S}},\boldsymbol{z_{\rm S}}\right\}$: $\boldsymbol{z_{\rm S}}\equiv\boldsymbol{z}$, where the $\boldsymbol{x_{\rm S}}$-axis is directed towards the position of a static and non-rotating observer at infinity $O_{\rm \infty}$, namely from $CM$ to $O'_{\rm \infty}$ (being the projection of $O_{\rm \infty}$ on $\mathcal{P}$). This RF is fixed in space and all quantities measured in it are labelled by a subscript S.
    
\end{itemize}

These RFs are related by the map $T:\mathcal{R}_{\rm CO}\rightarrow\mathcal{R}_{\rm S}$. It is determined by employing polar coordinates, the radius $r_{\rm S}(t)$ (coincident with the coordinate $x$) and polar angle $\theta_{\rm S}(t)$, where $t$ is the time. More explicitly, this transformation reads as 
    \begin{equation} \label{eq:T}
    \begin{cases}
    x_{\rm S}(t)=r_{\rm S}(t)\cos\theta_{\rm S}(t),\\
    y_{\rm S}(t)=r_{\rm S}(t)\sin\theta_{\rm S}(t),
    \end{cases}\qquad\Leftrightarrow\qquad r_{\rm S}(t)= x.
    \end{equation}

\subsection{Equations of motion}
\label{sec:EOMs}
Our model is governed by the gravitational quadrupole-orbit dynamics (see Sec. \ref{sec:G2P}) and the time-variation of the quadrupole moment (see Sec. \ref{sec:quadrupole_dynamics}). This differential problem can be solved if it is accompanied by the appropriate initial conditions (see Sec. \ref{sec:IC}). We stress again that, even though our model is fully based on the Applegate mechanism, the related dynamical equations of motion have never been written in the literature.

\subsubsection{Gravitational quadrupole-orbit coupling dynamics}
\label{sec:G2P}
The two bodies are influenced by their mutual gravitational attraction and the companion star's gravitational quadrupole-orbit coupling. The problem is first framed in $\mathcal{R_{\rm CO}}$, where the pulsar has coordinates $(-x_p,0,0)$, while the companion star $(x_c,0,0)$. 

The force acting on the pulsar is $F_p$, being the sum of the gravitational force and the quadrupole moment contribution $Q(t)$\footnote{The quadrupole moment is a tensor written in terms of the companion star's inertial tensor. In our hypotheses, we have $Q=Q_{xx}$ \citep[see discussion under Eq. (3) in][for details]{Applegate1992}.} from the companion star, namely \citep{Applegate1992}:
\begin{equation}\label{eq:pulsar}
m_p \ddot{x}_p=F_p\equiv-\partial_x\left(\frac{Gm_pm_c}{x}+\frac{3}{2}\frac{GQ(t)m_p}{x^3}\right).
\end{equation}

Instead, the force acting on the companion star is $F_c$, simply given by the gravitational force from the pulsar:
\begin{equation}\label{eq:compan}
m_c \ddot{x}_c=F_{c}\equiv \partial_x\left(\frac{Gm_pm_c}{x}\right).
\end{equation}

We define the relative coordinate system $x=x_c-x_p$ and the relative acceleration $\ddot{x}=\ddot{x}_c-\ddot{x}_p$ (the over dot stands for the derivative with respect to the time $t$), together with the total $M_{\rm TOT}=m_p+m_c$ and reduced $\mu=m_p m_c/M_{\rm TOT}$ masses. After appropriately manipulating Eqs. \eqref{eq:pulsar} and \eqref{eq:compan}, we obtain
\begin{equation} \label{eq:rel_eq}
\ddot{x}=-\frac{GM_{\rm TOT}}{x^2}-\frac{9}{2}\frac{GQ(t)}{x^4} .
\end{equation}

Changing RF through the transformation $T$, we rewrite the above dynamics in polar coordinates in $\mathcal{R}_{\rm S}$. Considering the angular component, we obtain the following equation of motion
\begin{equation}\label{eq:EoMD}
\frac{1}{r_{\rm S}(t)}\frac{\dd}{\dd t}\left[\mu\ r_{\rm S}(t)^2\ \dot{\theta}_{\rm S}(t)\right]=0,
\end{equation}
which immediately entails $\mu r_{\rm S}(t)^2\dot{\theta_{\rm S}}(t)=L$, where $L$ is the module of the conserved angular momentum $\textbf{L}$ of the system along the $\boldsymbol{z_{\rm S}}$-axis, which can be calculated using the initial conditions.

The problem is defined in the timeframe $[t_0,t_1]$. However, since the setting is invariant under time shifts, we can consider, without loss of generality, the following normalized interval $[0,1]$\footnote{We prefer to work in normalized units for developing the calculations, since this is advantageous during the fitting procedure.}. The map connecting $[t_0,t_1]$ with $[0,1]$ is given by
\begin{equation} \label{eq:time-transformation}
t\in[t_0,t_1]\to \frac{t-t_0}{t_1-t_0}\in [0,1].    
\end{equation}
We stress that $T_0$ is the reference time, usually coincident with the transition to the orbit periastron\footnote{For circular orbits, $T_0$ is the ascending node's epoch passage.}. This is the moment where we extract the parameters and would correspond to the initial time $T_0^*=(T_0-t_0)/(t_1-t_0)$. However, the observations begin at $t_0 < T_0$. Therefore, it is reasonable to set the initial conditions at $T_0$ and extend our solutions back to the earlier time, $t_0$.

The angle $\theta_{\rm S}(t)$ can be calculated through the formula $\theta_{\rm S}(t)=\omega(t) (t-T_0^*)$, where $\omega(t)$ is the angular frequency. Since the system is very tight, the bodies move on quasi-circular orbits with relative angular velocity \citep[see Eq. (5) in][]{Applegate1992}
\begin{equation} \label{eq:v_theta}
v_\theta(t)=\sqrt{\frac{GM_{\rm TOT}}{r_{\rm S}(t)}\left(1+\frac{9}{2}\frac{Q(t)}{m_c r^2_{\rm S}(t)}\right)}\,. 
\end{equation}
Therefore, the angular frequency $\omega(t)$ can be estimated applying the formula for (quasi-)circular motion \citep{Applegate1992}
\begin{equation} \label{eq:omega1}
\omega(t)=\frac{v_\theta(t)}{r_{\rm S}(t)}.
\end{equation}
It is important to note that $\dot{\theta}_{\rm S}(t)\neq \omega(t)$, since
\begin{equation}
\dot{\theta}_{\rm S}(t)=\omega(t)+(t-T_0^*)\dot{\omega}(t).
\end{equation}
However, at the beginning (i.e., for $t=T_0^*$) we have $\dot{\theta}_{\rm S}(T_0^*)=\omega(T_0^*)$.

Instead, for the radial component, we obtain 
\begin{equation} \label{eq:EoM1}
\mu \ddot{r}_{\rm S}(t)=\frac{L^2}{\mu r_{\rm S}^3(t)}-\frac{Gm_pm_c}{r^2_{\rm S}(t)}-\frac{9}{2}\frac{GQ(t) m_p}{r^4_{\rm S}(t)},
\end{equation}
where the change of RF can be seen by the appearance of the centrifugal force (first term on the right hand side). This dynamical system is composed by a second-order ordinary differential equation \eqref{eq:EoM1}, which should be complemented by the dynamics of $Q(t)$\footnote{It is important to note that being $Q$ directed along the $x$-axis, thanks to the transformation \eqref{eq:T}, we have that $Q(t)$ points along $r_{\rm S}(t)$.}, which is disclosed in the next section. 

\subsubsection{Quadrupole dynamics}
\label{sec:quadrupole_dynamics}
The Applegate mechanism foresees that there are two kinds of deformations due to the magnetic activity, which can be classified into: (1) \emph{distortions}, which modify the hydrostatic equilibrium in the deformed configuration; (2) \emph{transitions}, which cause changes from one fluid hydrostatic configuration to another.

\citet{Applegate1992} explains that distortions are not astrophysically relevant, because the weak magnetic fields ($\sim10^5-10^6$ G) cannot supply enough energy for the star deformations to reproduce the orbital period modulation timescales (see after Eq. (23), for a more detailed discussion). Therefore, we consider modifications due to the transitions, where the dynamics of a rotating star strongly depends on the matter distribution within it and its angular momentum $J$, which influences the quadrupole moment variations in agreement with the observed timescales. 

A fundamental role is played by its external layers, which contribute to spin-up the star, making it more oblate \citep[i.e., enhancing its quadrupole moment;][]{Applegate1992}. Therefore, the magnetic activity permits to arise and develop a torque, which acts on the spin of the companion star to extract angular momentum \citep{Applegate1992,Applegate1994}. From Eq. (26) in \citet{Applegate1992}, the companion star's quadrupole moment changes according to the following formula:
\begin{equation} \label{eq:Q}
\dot{Q}=\frac{1}{3}\left(\frac{\Omega R_c^3}{Gm_c}\right)\dot{J}, 
\end{equation}
where $\Omega$ is the angular velocity of the outer layers, which can be reasonably calculated through the Keplerian angular velocity   
\begin{equation} \label{eq:omega2}
\Omega=\sqrt{\frac{Gm_c}{{R_c}^3}}.    
\end{equation}
The variation of $J$ is caused by the spin torque, because the irradiation driven wind from the pulsar generates a ram pressure contributing to the spin-up of the companion star. Employing Eq. (27) in \citet{Applegate1992}, the time-variation of $J$ is\footnote{It is important to note that $\Delta P(t)=P(t)-P_0$ with $P_0$ being a constant. Therefore, $\dd\Delta P/\dd t=\dot{P}(t)$. We use both notations in the paper.}
\begin{equation} \label{eq:dotJ}
\dot{J}=-\frac{Gm_c^2}{6\pi R_c}\left(\frac{r_{\rm S}(t)}{R_c}\right)^2\frac{\dd\Delta P}{\dd t},
\end{equation}
where $\Delta P(t)$ is also known as \emph{orbital period modulations}. Therefore, the resulting differential equation ruling the dynamics of $Q(t)$ is obtained by substituting Eq. \eqref{eq:dotJ} into Eq. \eqref{eq:Q}, namely
\begin{equation}\label{eq:EoM2}
    \dot{Q}(t)=- \dfrac{m_c \Omega 
 r_{\rm S}^2(t)}{18 \pi} \dfrac{\dd \Delta P}{\dd t}.
\end{equation}

\subsubsection{Initial conditions}
\label{sec:IC}
Our model is governed by a system of ordinary differential equations \eqref{eq:EoM1} and \eqref{eq:EoM2}. It must be complemented by the initial conditions at the time $t=T_0^*$ to find a unique solution, which are
\begin{equation} \label{eq:ICN}
\dot{r}_{\rm S}(T_0^*)=\frac{2}{3}\frac{\dot{P}(T_0^*)}{P(T_0^*)}a,\ r_{\rm S}(T_0^*)=a,\ \theta_{\rm S}(T_0^*)=0, \ Q(T_0^*)=Q_0,    
\end{equation}
where $a$ is the initial separation between pulsar and companion star and $Q_0=0.1m_c R_c^2/3$ the initial quadrupole \cite[cf. Eq. (25) in][]{Applegate1992}. The radial velocity is determined employing Kepler's third law in its differential form, as the two bodies' orbits are Keplerian at every moment in time. The conditions \eqref{eq:ICN} imply (cf. Eqs. \eqref{eq:v_theta} and \eqref{eq:omega1})
\begin{equation} \label{eq:VA}
v_\theta(T_0^*)=\sqrt{\frac{GM_{\rm TOT}}{a}\left(1+\frac{9}{2}\frac{Q_0}{m_c a^2}\right)},\qquad \omega(T_0^*)=\frac{v_\theta(T_0^*)}{a}. 
\end{equation}

\subsection{Physical observables}
\label{sec:observables}
This section provides some physical observables related to spider systems. To achieve this objective, we must first solve Eqs. \eqref{eq:EoM1} -- \eqref{eq:EoM2}, as we need $r_{\rm S}(t)$ and $Q(t)$, which are already fundamental physical quantities. We focus on the following quantities: orbits pertaining to the two bodies (see Sec. \ref{sec:orbit}), orbital period (see Sec. \ref{sec:orbper}), magnetic field intensity (see Sec. \ref{sec:magnetic}), luminosity variability (see Sec. \ref{sec:LV}).

\subsubsection{Two body orbits}
\label{sec:orbit}
An important feature of a binary system is to understand the evolution of the orbit. Since the two bodies are tidally locked and synchronized, we need to determine the separation $r_{\rm S}(t)$ and the quadrupole moment $Q(t)$ to calculate $\omega(t)$ (cf. Eq. \eqref{eq:omega1}), which in turn provides the evolution of the polar angle $\theta_{\rm S}(t)$. Therefore, passing in cartesian coordinates, the orbit is obtained by plotting
\begin{equation}
(r_{\rm S}(t)\cos\theta_{\rm S}(t),r_{\rm S}(t)\sin\theta_{\rm S}(t)),\qquad t\in[0,1]    
\end{equation}
where $t$ is given by Eq. (\ref{eq:time-transformation}).

\subsubsection{Orbital period}
\label{sec:orbper}
The orbital period $P(t)$ can be estimated in two ways: (1) once we solve the system, we determine the orbit and we then calculate it ; (2) it can be obtained \emph{a-priori} by fitting the observational data. In this work, we use the latter approach, as we aim to reconstruct the dynamics of the physical observables.

In the \emph{a-priori approach} we first calculate $\Delta P(t)=P(t)-P_0$, where $ P_0$ is estimated via Kepler's third law, namely
\begin{equation} \label{eq:unbert-period}
P_0=\sqrt{\frac{4\pi^2 a^3}{G M_{\rm TOT}}}.
\end{equation}
The formula used to fit $\Delta P (t)$ (generally expressed in seconds) is
\begin{equation}\label{eq:delta_P}
    \Delta P(t) =   \dfrac{1}{g(t)} - \dfrac{1}{f_0},
\end{equation}
where $f_0=1/P_0$ and
\begin{align}
    g(t)&\simeq f_0 + f_1 (t-T_0) + \dfrac{1}{2} f_2 (t-T_0)^2 + ... + \dfrac{1}{n!}f_n(t-T_0)^n\notag\\
    &=\sum\limits_{i=0}^n \frac{1}{i!}f_i(t-T_0)^i.
\end{align}
The fitting procedure is performed in the interval $[t_0,t_1]$, expressed in MJD. Furthermore, the coefficients $\{f_i\}_{i=0}^n$ represent the higher-order frequency derivatives and $n$ is the order of terms involved in the fit. The parameters $\{f_i\}_{i=0}^n$ and $T_0$ are obtained as a timing solution of a BTX model provided by the \texttt{TEMPO} or \texttt{TEMPO2} software. The value of $n$ changes from a pulsar to another, since it is the number of the reliably-measured time derivatives of the orbital period, and depends on various aspects, such as: the time interval covered by the radio observations, the typical uncertainty on the measured pulses' time of arrival, the r.m.s of the timing residuals, and more \citep[see e.g.,][]{Ridolfi2016,Freire2017}. We also note that $\dot{P}(T_0^*)=-f_1/f_0^2$, useful for computing $\dot{r}(T_0^*)$ in Eq. \eqref{eq:ICN}.

For what follows, it is useful to define the constant quantity
\begin{equation} \label{eq:deltaPoverP}
\mathcal{A}=\langle \frac{\Delta P(t)}{P(t)} \rangle_{[t_0,t_1]},    
\end{equation}
where $\langle \cdot \rangle_{[t_0,t_1]}$ is the time-average in the interval $[t_0,t_1]$.

\subsubsection{Magnetic field intensity}
\label{sec:magnetic}
In the Applegate mechanism, the magnetic energy is the main source of support to provide the necessary torque for the exchange of angular momentum between the shells of the companion star (and consequently changes in the quadrupole moment). In this scenario, the magnetic field does not decay in rapid times, because the so far observations have not detected orbital period variations over short timescales. This phenomenon represents an indirect probe for the internal magnetic field dynamics within active stars. Therefore, spider pulsars are natural laboratories and privileged systems to investigate the magnetic dynamics in low-mass stars, since this topic is still matter of discussion.  

The variation of the subsurface (toroidal) magnetic field intensity with respect to that of the unperturbed (i.e., $Q(t)=0$) configuration can be estimated through the following formula \citep[see Eq. (33) in][for further details]{Applegate1992} 
\begin{equation} \label{eq:magfield}
\Delta B(t) \sim\rm{sgn}(\Delta P(t)) \sqrt{10\frac{Gm_c^2}{R_c^4}\left(\frac{r_{\rm S}(t)}{R_c}\right)^2\frac{|\Delta P(t)|}{P_{ma}}}\, ,
\end{equation}
where $P_{\rm ma}$ is the magnetic period, being specific for each spider system. It is computed by searching for the maxima (or minima) in the $\Delta P(t)$ profile, whose average distances allows us to extract a mean value. We stress that Eq. \eqref{eq:magfield}, as derived by \citet{Applegate1992}, provides just an estimate of the magnetic field intensity. In the original formula, $\Delta P$ is interpreted as a (constant) positive quantity. Since in our case $\Delta P(t)$ assumes both positive and negative values, as well as zero, it is more appropriate to insert the absolute value to derive a realistic result. In addition, inspired by the $\Delta P(t)$ profile, it is more opportune to calculate (through Eq. \eqref{eq:magfield}) the variations $\Delta B(t)$, rather than $B(t)$. We stress $\Delta B(t)=B(t)-B_0$, where $B(t)$ is due to the quadrupole variations, whereas $B_0$ stays for the unperturbed case. In order to guarantee regular behaviours and attain negative values, we have added the function sign of the $\Delta P(t)$ outside the square root. We make the function continuous, but it is not differentiable in the points crossing the zero line. This is the best we can achieve, as we are ignorant about the magnetic activity occurring inside the companion star.

We consider a rough estimate of the magnetic field $B_0$ proposed by Applegate, which is obtained by substituting in Eq. \eqref{eq:magfield}, $r_{\rm S}(t)$ with its initial value $a$ and $\Delta P(t)$ with the trend $\mathcal{A}P_0$ (cf. Eq. \eqref{eq:deltaPoverP}). Therefore, performing these calculations, we have
\begin{equation} \label{eq:magfield_App}
B_0 \sim \rm{sgn}(\mathcal{A}) \sqrt{10\frac{Gm_c^2}{R_c^4}\left(\frac{a}{R_c}\right)^2\frac{\mathcal{A}P_0}{P_{ma}}}\, .
\end{equation}

\subsubsection{Luminosity variability}
\label{sec:LV}
Another non-uniform periodic mechanism is related to the \emph{physics of the luminous variability} inside the active star. This cycle is divided into two phases \citep{Applegate1987,Applegate1992,Applegate1994}: 
\begin{itemize}
    \item[$(i)$] the angular momentum transfer leads to an enhancement of the kinetic energy, because it is spent to power the differential rotation between the inner part and the outer layer of the active star. This activity leads the star to spin-up at the expenses to lower the luminosity;
    \item[$(ii)$] when the angular momentum transfer decreases, the active star rotates as a solid body. This causes the star to spin-down, with a consequent increment in its luminous intensity.
\end{itemize}
The alternation of the two cycles gives rise to the observed luminous modulations. In addition, in the entire process there is also a weak dissipation, because the orbital period tends to weakly decrease with time. However, this dissipative effect will be not taken into account in this model due to its negligible contributions.

The energy emitted by the companion star can be estimated through \citep[see Eq. (28) in][for details]{Applegate1992}
\begin{equation}\label{eq:L}
\Delta E=\Omega_{\rm dr}\dot{J}+\frac{\dot{J}^2}{I_s}\Delta t,   
\end{equation}
where $\Omega_{\rm dr}=\Omega(1-\eta)$ with efficiency $\eta=0.66$ \citep[see Fig. 1 in
][]{Yoshida2019} and $I_s$ represents the moment of inertia of the outer layer considered as a shell, given by \citep{Applegate1992}
\begin{equation}\label{eq:MI}
I_s=\frac{2}{3}M_s R_c^2,    
\end{equation}
where $M_s\approx 0.1 m_c$ is the outer layer mass \citep{Applegate1992}. Therefore, the luminosity modulation can be easily calculated as 
\begin{equation} \label{eq:luimnosity}
\Delta\mathscr{L}(t)=\pi\frac{\Delta E}{P_{\rm ma}},    
\end{equation}
where $\Delta\mathscr{L}(t)$ represents the difference between the luminosity due to the quadrupole variations and the constant luminosity of the star without altering its shape. This formula permits to track the companion star's luminosity with respect to the unperturbed configuration during the time evolution, and to monitor how it changes in terms of the orbital period modulations. 

Finally, we provide an estimate of the unperturbed luminosity, expressed by the following formula (using Eq. \eqref{eq:luimnosity}, where we substitute $r_{\rm S}(t)$ with $a$ and $\Delta P(t)$ with $\mathcal{A}P_0$)
\begin{equation}\label{eq:luminsity_fix}
\mathscr{L}_0=\frac{1}{9}\Omega \frac{G m_c^2}{R_c} \left(\frac{a}{R_c}\right)^2   \frac{\mathcal{A} P_0}{P_{\rm mod}}.
\end{equation}

\section{Methodology}
\label{sec:approx_meth}
The spider system dynamics is described by two coupled non-linear differential equations \eqref{eq:EoM1} -- \eqref{eq:EoM2}, where the analytical solution is too difficult to be determined and therefore, numerical routines must be exploited. We note that if we follow the a-posteriori approach (see Sec. \ref{sec:orbper}), we must have a precise temporal trend for $\dd\Delta P/\dd t$, which is related to the subsurface magnetic field of the active star. This is a very demanding task for several reasons \citep[see][and references therein]{simul_2006,simul_2008}: (1) theoretical uncertainties about the microphysics inside low-mass stars; (2) high computational cost toperform magnetohydrodynamic (MHD) simulations; (3) existence of several models, based on different simplifications and hypotheses. 

In order to avoid the aforementioned issues, we reckon on the a-priori approach (see Sec. \ref{sec:orbper}), founded on having the function $\Delta P(t)$ by fitting the observational data. However, even after substituting the orbital period modulations \eqref{eq:delta_P} in the equations of motion, the ensuing dynamics is still too cumbersome to be analytically integrated. Therefore, if we want to avoid numerical integrations, some approximation schemes must be employed. 

We present a methodology to derive an approximate analytical solution of Eqs. \eqref{eq:EoM1} and \eqref{eq:EoM2} in Sec. \ref{sec:succ_stra}, about which we comment on the approximation accuracy with respect to the original (unaffected) equations in Sec. \ref{sec:accuracy}. We conclude by specifying the inputs and outputs of our model in Sec. \ref{sec:code}.

\subsection{Approximation strategy}
\label{sec:succ_stra}
Analysing better the problem, we should bare in mind that the motion is quasi-circular. Therefore, we can assume the validity of Kepler's third law in each point. Differentiating it, we obtain, at the first order in $\Delta P$ and $\Delta r_{\rm S}$:
\begin{equation}\label{eq:approx}
 \dfrac{\Delta P}{P} = \frac{3}{2} \dfrac{ \Delta r_{\rm S} }{r_{\rm S}}.
\end{equation}
Substituting the following relations:
\begin{equation}\label{eq:definitions}
    P(t)= P_0 + \Delta P (t),\qquad r_{\rm S}(t)= a + \Delta r_{\rm S} (t)
\end{equation}
into Eq. \eqref{eq:approx} and neglecting second order terms, we have
\begin{equation}\label{delta_r}
    \Delta r_{\rm S} (t) = \frac{2}{3} \dfrac{a  \Delta P (t)}{2 P_0 },
\end{equation}
which completely determines the function $r_{\rm S}(t)$ in terms of the orbital period modulations $\Delta P(t)$, taken from the observations. 

Equation \eqref{eq:EoM2} is the only differential relation left, which must be solved numerically. However, we can avoid this last integration by sampling $r_{\rm S}(t)$ and $\Delta P(t)$ functions at several points. Then, we fit them with high-accurate polynomials with $n+1$ coefficients, using the same $n$ of the \texttt{TEMPO} coefficients $\{f_i\}_{i=0}^n$, because this allows a drastic reduction of the approximation errors. The fitting procedure occurs in the interval $[0,1]$, since this gives more accurate results. Using the transformation \eqref{eq:time-transformation} we can  then map the ensuing polynomial in the interval $[t_0,t_1]$.

In this case, Eq. \eqref{eq:Q} can be analytically written as
\begin{equation} \label{eq:func_form_Q}
Q(t)=-\frac{m_c\Omega}{18\pi}\sum_{i=0}^{m} a_i \frac{t^{i+1}}{i+1}+Q_0,   
\end{equation}
where the coefficients $a_i$ are real numbers, which can be explicitly calculated when we have the polynomials of $r_{\rm S}(t)$ and $\Delta P(t)$. The integer $m$ depends on the final polynomial order obtained by multiplying the two aforementioned polynomials.

\subsection{Approximation accuracy}
\label{sec:accuracy}
To check the reliability of the result we have found, we need to compare our approximation with the numerical solution of Eqs. \eqref{eq:EoM1} and \eqref{eq:EoM2}. To this purpose, we use \texttt{Mathematica 13.1} and \texttt{Python 3} to confidently validate our calculations.

In \texttt{Mathematica 13.1}, we employ the function \texttt{NDSolve} and exploiting the methods \texttt{StiffnessSwitching} and \texttt{ExplicitRungeKutta}, selecting a precision and accuracy of 10, and a maximum step size of $10^{-6}$ in the interval $[0,1]$. Then, we plot $r_{\rm S}$ with several points ($\sim800$), while $Q(t)$ can be displayed by employing much fewer points ($\sim100$).

In \texttt{Python 3}, we use the integration routine \texttt{dop853}, being the \emph{Dormand-Prince algorithm} implemented within the class of explicit Runge-Kutta methods of eight order \cite{DORMAND1980,Press2002}. We select absolute $10^{-20}$ and relative $10^{-10}$ tolerances within $[0,1]$ with a step-size of $10^{-7}$. The two approaches are in agreement, since they give the same results.

The approximate radius differs from the numerical one, because the latter contains mild oscillations (since the orbit is quasi-circular) and an overall modulation on the time span $[0,1]$; whereas the former features only the modulation on $[0,1]$ (since the orbit is considered circular). Instead, the quadrupole moment between the two approaches coincides with mean relative errors (MREs) $<10^{-5}\%$. Therefore, our solution is consistent.

\subsection{Inputs and outputs of the model}
\label{sec:code}
Our approach relies on the analytical (high-accurate approximate) formulae of $r_{\rm S}(t)$ and $Q(t)$. This permits to fast compute the evolution of the associated physical observables reported in Sec. \ref{sec:observables}. Our methodology is also flexible, because it can be used in an opposite manner. Indeed, knowing the trend of some physical variables (e.g., the luminosity), we could extract the orbital period modulations and then determine all the other quantities.

The input values of our model are:
\begin{equation} \label{eq:input}
\Big\{m_p,m_c,R_c,P_{\rm ma},f_0,\dots,f_n,T_0\Big\},    
\end{equation}
where $m_p$ can be set equal to the value of a standard NS (i.e., $m_p=1.4$\msun) and if we know the spider class, we can assign an average value of $m_c$ and $R_c$. The initial orbital separation $a$ can be calculated via the third Kepler's law (cf. Eq. \eqref{eq:unbert-period}), knowing $P_0=1/f_0, m_p,m_c$. Depending on the goal, we generally have $n+6$ input parameters, which can eventually be lowered to $n+3$.

The output parameters of our model are: 
\begin{equation} \label{eq:output}
\Big\{r_{\rm S}(t),\theta_{\rm S}(t),Q(t),\Delta B(t),\Delta\mathscr{L}(t)\Big\}.   
\end{equation}

\section{Results}
\label{sec:results}
\begin{figure*}[ht!]
    \centering
    \hbox{
    \includegraphics[scale=0.245]{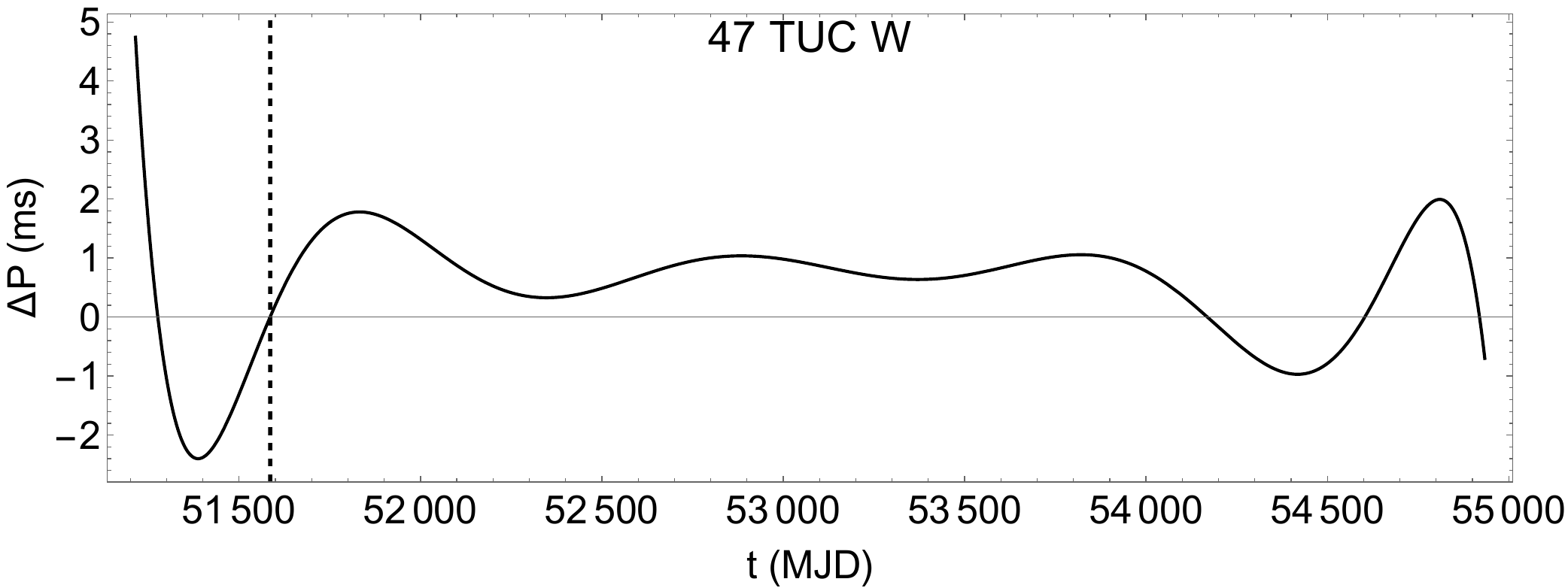}\hspace{0.65cm}
     \includegraphics[scale=0.245]{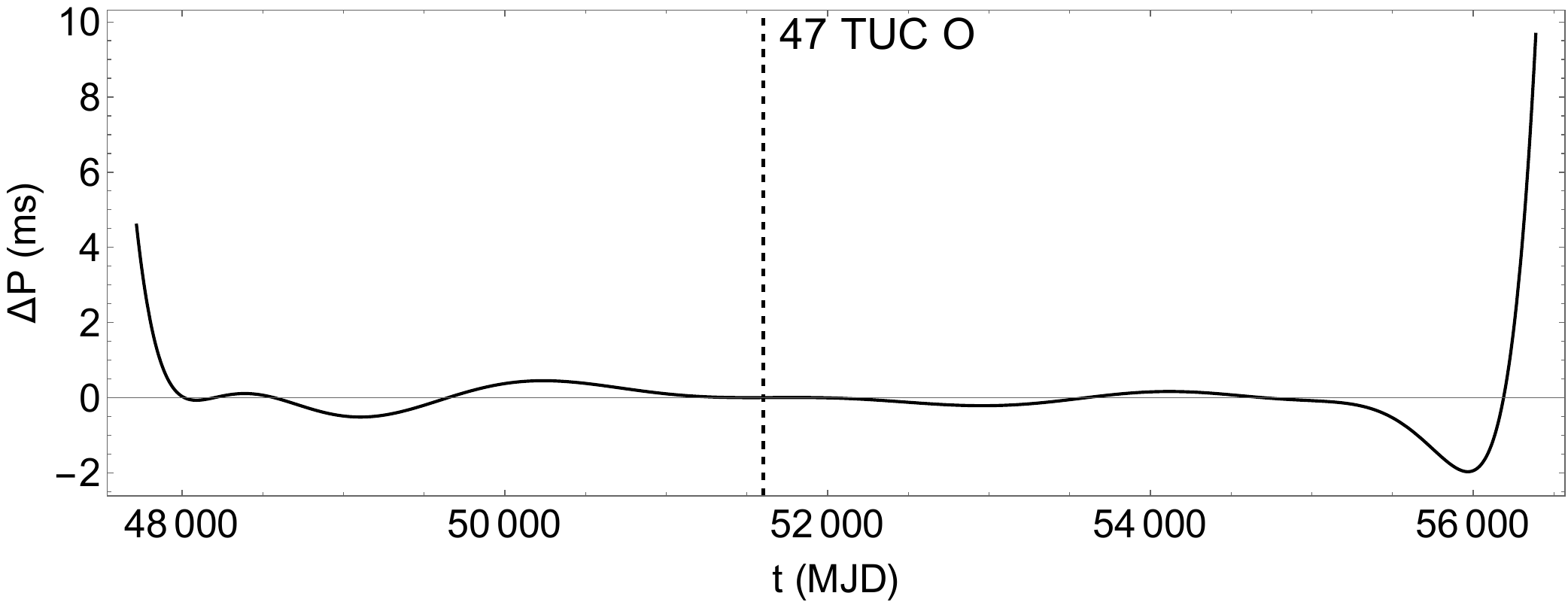}
    }
    \caption{Orbital period modulations of 47 Tuc W and 47 Tuc O, obtained from the fitting of the observational data (cf. Table \ref{tab:Table1}). The vertical dashed line marks the position of $T_0$.}
    \label{fig:Fig3}
\end{figure*}

As an application of our model, we consider the spider systems: 47~Tuc~W (redback) and  47~Tuc~O (black widow). The input parameters and the \texttt{TEMPO} coefficients of these two physical systems are reported in Table \ref{tab:Table1}. In Fig. \ref{fig:Fig3} we display the orbital period modulations of these two sources \citep[see][for more details]{Ridolfi2016,Freire2017}. The polynomial approximations described in Sec. \ref{sec:succ_stra} are reported in Appendix \ref{appendix}. In this section, we first compare the results from the two sources in Sec. \ref{sec:comparison} and then in Sec. \ref{sec:parameter-determination} we analyse how we determine the associated parameters, once we detect a spider source.

\subsection{Comparison between 47 Tuc W and 47 Tuc O}
\label{sec:comparison}
\begin{figure*}[h!]
    \centering
    \hbox{
    \includegraphics[scale=0.245]{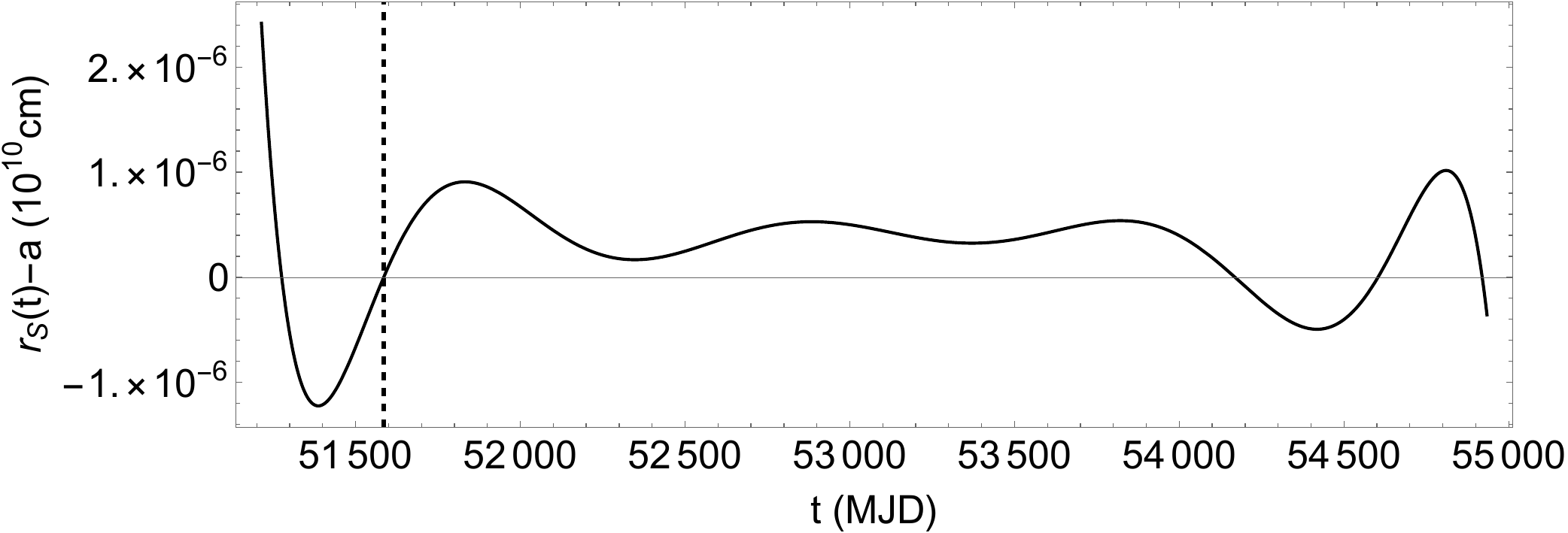}\hspace{0.65cm}
    \includegraphics[scale=0.245]{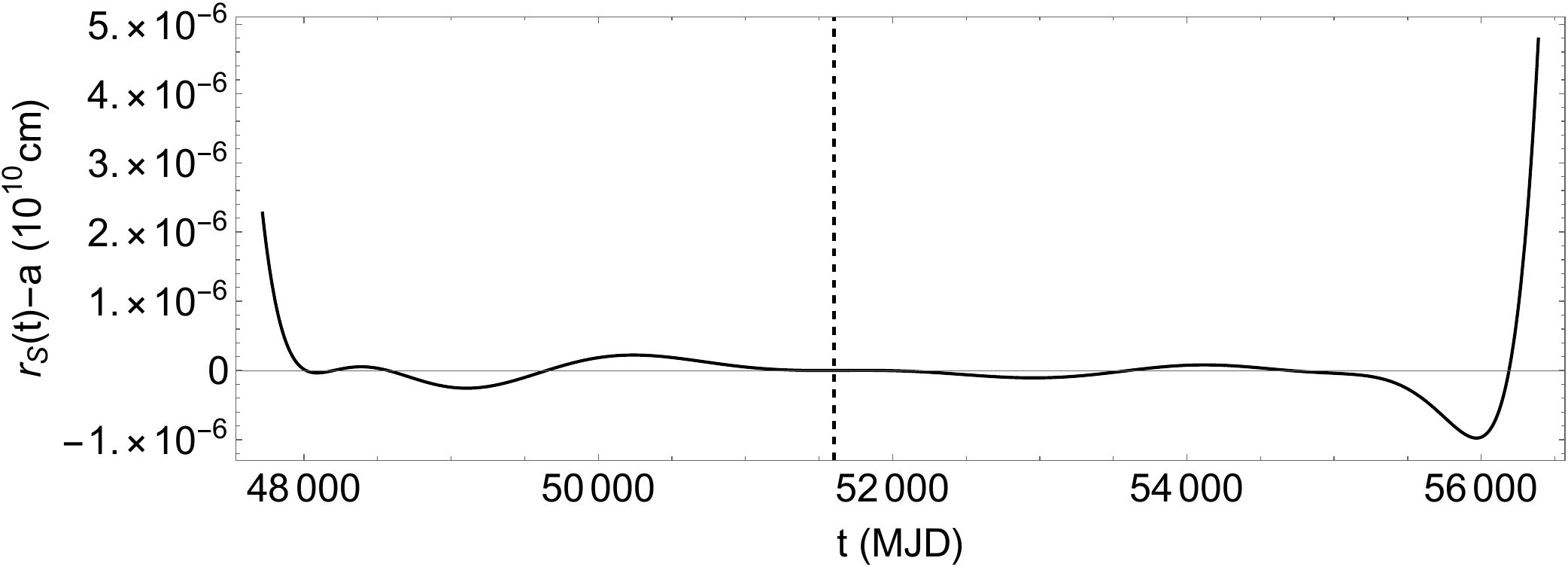}}
    \caption{Discrepancy of the actual separation among the bodies with respect to the initial datum. The vertical dashed line marks $T_0$. The order of the sources is placed as in Fig. \ref{fig:Fig3}.}
    \label{fig:radius}
\end{figure*}   

\begin{figure*}[h!]
    \centering
    \hbox{
    \includegraphics[scale=0.245]{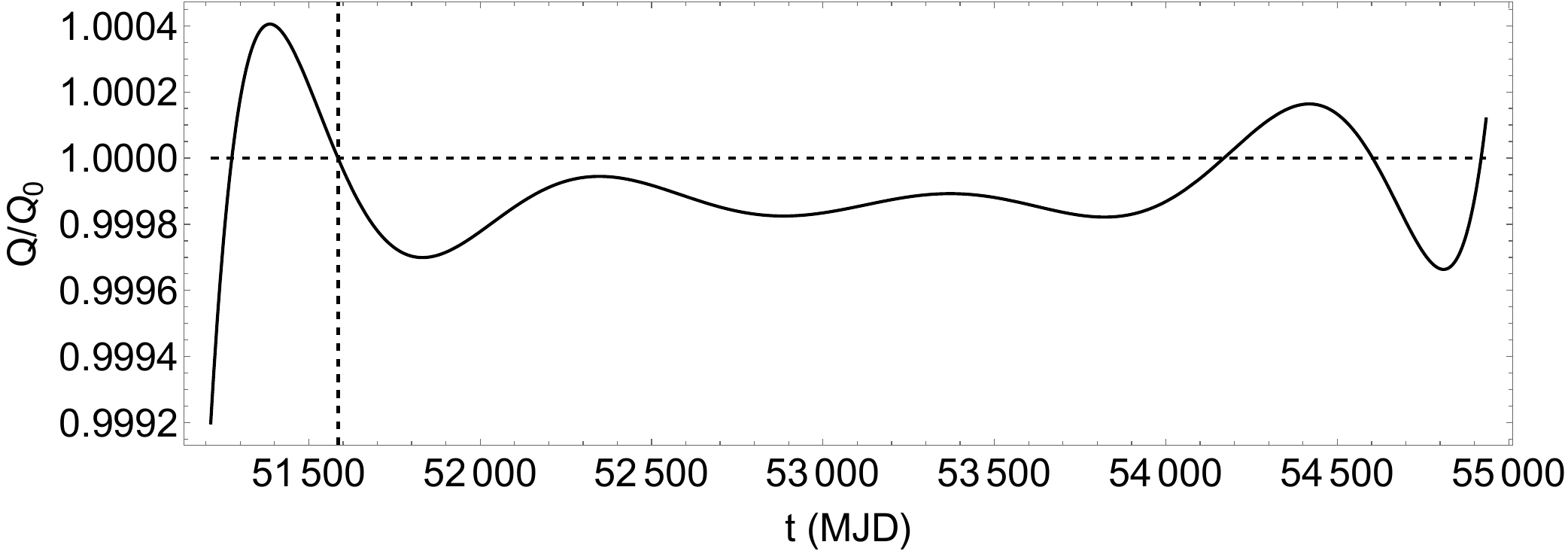}\hspace{0.65cm}
    \includegraphics[scale=0.245]{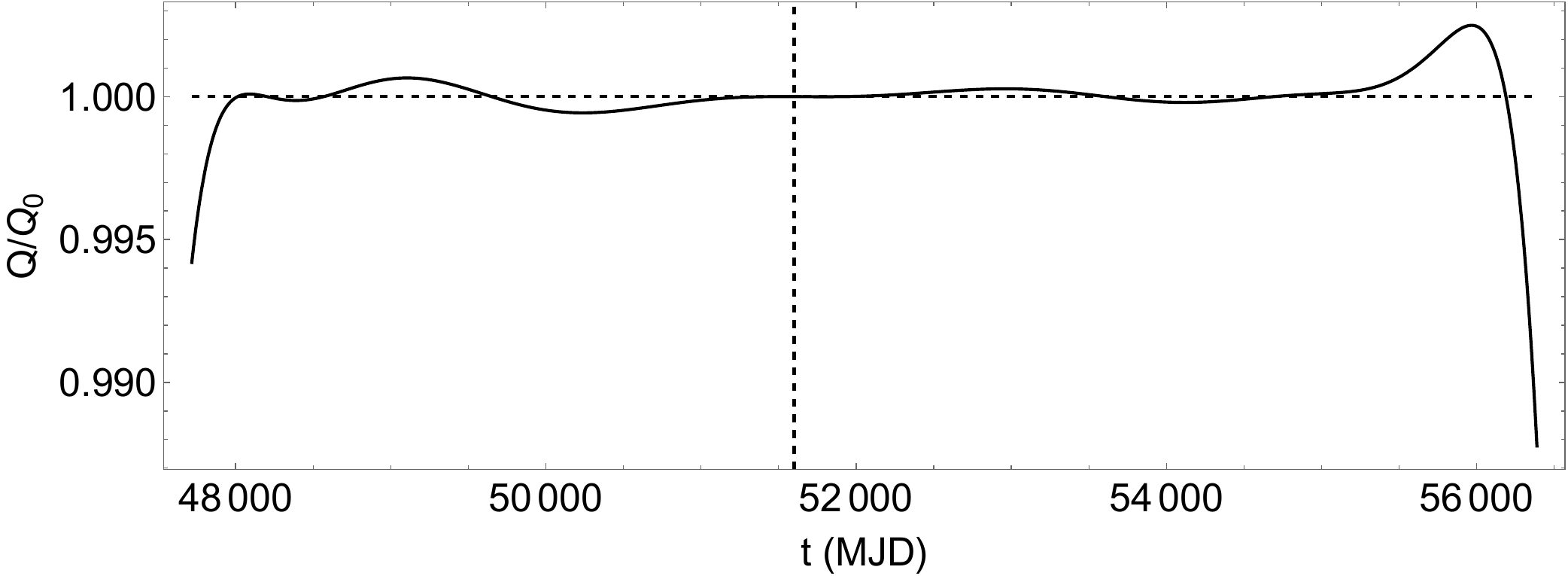}}
    \caption{Evolution of the quadrupole moment $Q(t)/Q_0$, where the horizontal dashed line is set at 1 and the vertical dashed line marks $T_0$. The order of the sources is placed as in Fig. \ref{fig:Fig3}.}
    \label{fig:quadrupole}
\end{figure*}    

\begin{figure*}[h!]
    \centering
    \hbox{
    \includegraphics[scale=0.245]{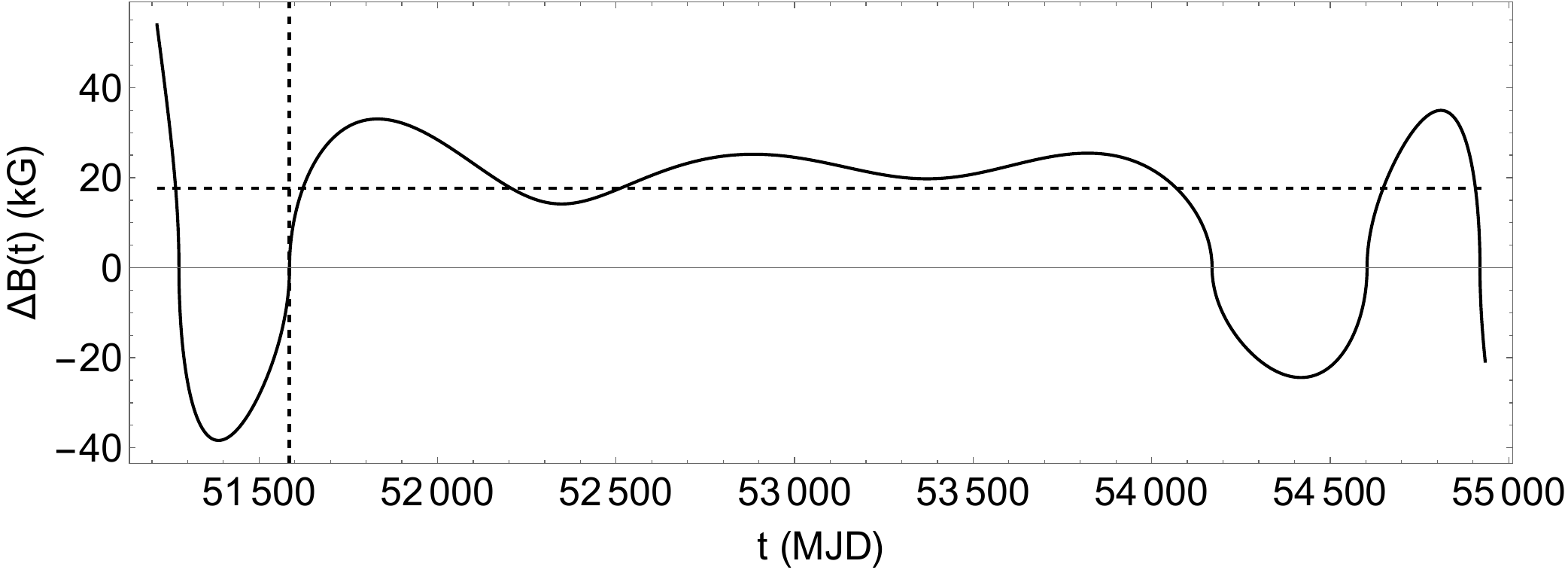}\hspace{0.65cm}
    \includegraphics[scale=0.245]{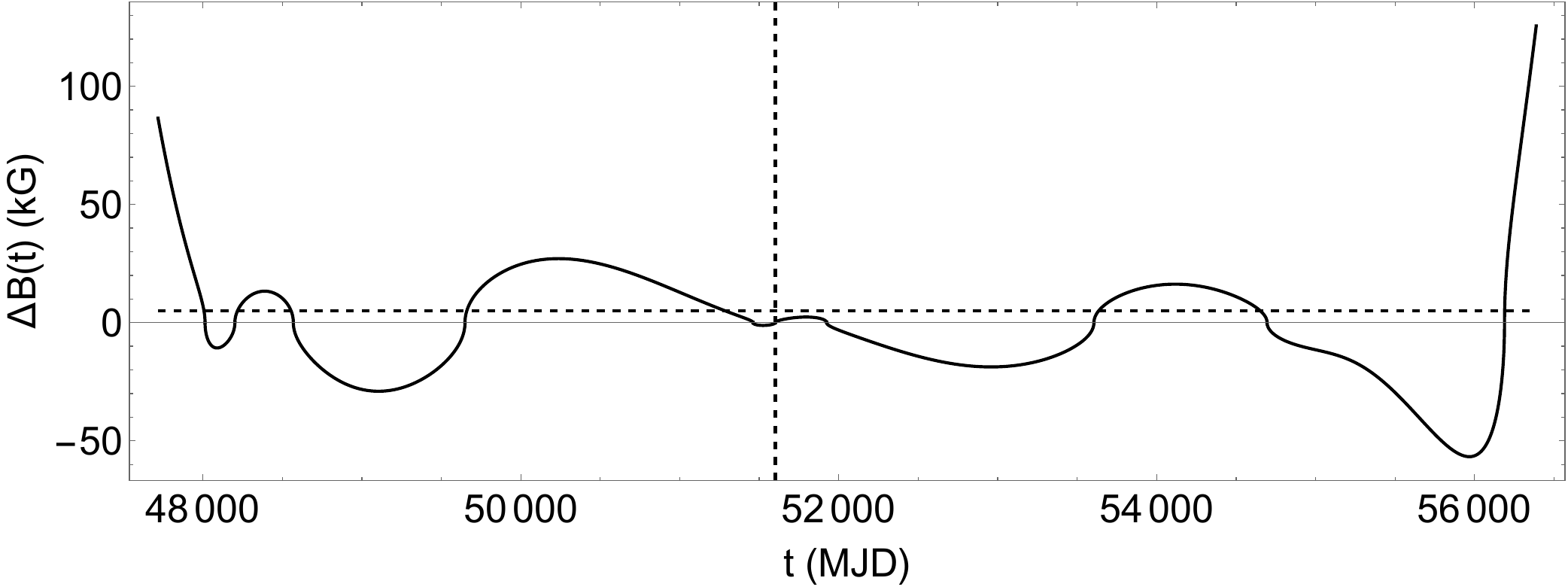}}
    \caption{Variation of magnetic field intensity $\Delta B(t)$. The horizontal dashed line is $B_0$ (cf. Eq. \eqref{eq:magfield_App}), whereas the vertical dashed line marks $T_0$. The order of the sources is placed as in Fig. \ref{fig:Fig3}.}
    \label{fig:magnetic}
\end{figure*}    

\begin{figure*}[h!]
    \centering
    \hbox{
    \includegraphics[scale=0.245]{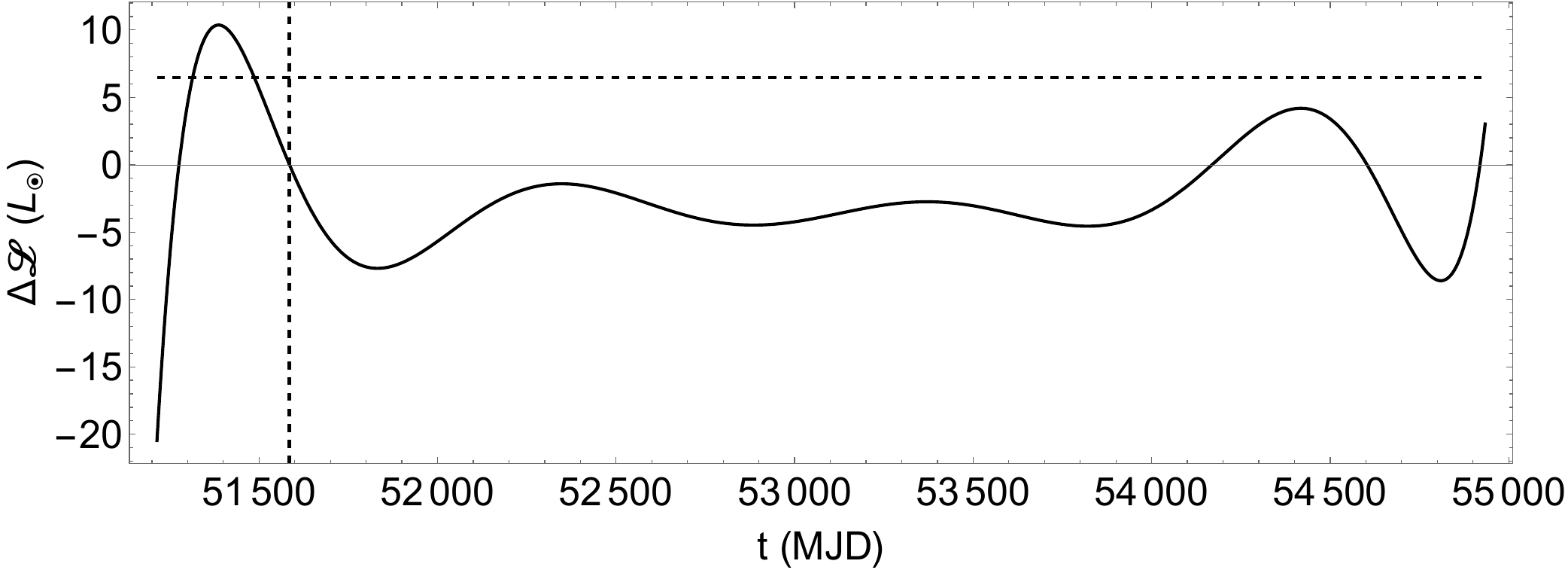}\hspace{0.65cm}
    \includegraphics[scale=0.245]{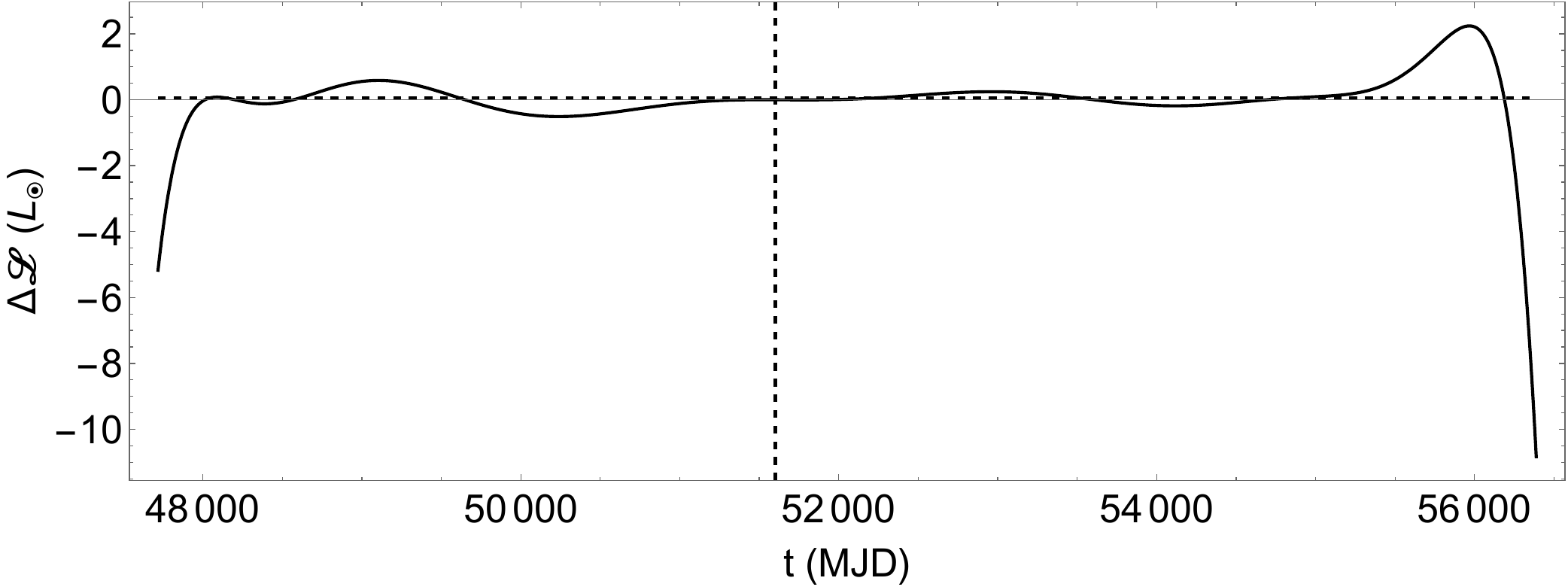}}
    \caption{Variation of luminosity $\Delta\mathscr{L}(t)$. The horizontal dashed line is $\mathscr{L}_0$ (cf. Eq. \eqref{eq:luminsity_fix}), whereas the vertical dashed line marks $T_0$. The order of the sources is placed as in Fig. \ref{fig:Fig3}.}
    \label{fig:luminosity}
\end{figure*}    

We analyse 47 Tuc W and 47 Tuc O's dynamics via our model through the related physical variables' profiles \eqref{eq:output}, highlighting common features (see Sec. \ref{sec:analogies}) and diversities (see Sec. \ref{sec:diversities}).

\subsubsection{Analogies}
\label{sec:analogies}
The radius follows the same trend of the orbital period modulations (cf. Eq. \eqref{eq:definitions} and \eqref{delta_r}), but with very mild oscillations, see Fig. \ref{fig:radius}. This feature transmits also to the orbits, which, besides to be quasi-circular, admit a very narrow advancement with respect to the long time baseline. Instead, the quadrupole moment behaves in the opposite way (see Fig. \ref{fig:quadrupole}), because the orbit shrinks (enlarges) as the quadrupole increases (decreases).

The magnetic field variability, compared to other observables, features a very oscillating trend scanned by the magnetic activity period $P_{\rm ma}$, see Fig. \ref{fig:magnetic}. The variable magnetic field $B(t)$ crosses the zero line at the points where $\Delta P(t)$ nullifies, corresponding to moments when the magnetic field matches the initial star's intensity. Since the provided formula is a rough estimate of the subsurface magnetic field strength (with an offset value $B_0$), we anticipate that MHD simulations could offer a similar but more detailed and accurate representation.

Finally, the luminosity is ruled by the quadrupole moment (cf. Eq. \eqref{eq:luimnosity}), where the zero points coincide with the luminosity of the star initially observed, see Fig. \ref{fig:luminosity}. For the calculation of the magnetic field and luminosity we use their original formulae (cf. Eqs. \eqref{eq:magfield} and \eqref{eq:luimnosity}) without any approximation for the radius $r_{\rm S}(t)$. We express the luminosity variability in solar luminosity units, corresponding to $\mathscr{L}_\odot=3.83\times10^{33}$ erg/s. We note that the results we have found are in agreement with the estimates reported in \citet{Applegate1992}, namely $\langle \Delta\mathscr{L}(t)\rangle_{[t_0,t_1]}/\mathscr{L}_0\sim 0.3$. 

\subsubsection{Differences}
\label{sec:diversities}
The described similarities are mainly due to the underlying equal mathematical structure, while the discrepancies arise from the different physical nature of the two spider classes. 

Observationally, the orbits associated with 47 Tuc W appear to be tighter than those of 47 Tuc O (see Fig. \ref{fig:radius}). This trend is consistent with theoretical models of spider pulsars, which suggest that the pulsar wind significantly impacts the companion’s structure and mass-loss process \citep{Chen2013,Wang2021,Conrad_Burton_2023}. The high-energy photons emitted by the pulsar deposit energy in the companion’s outer layers, altering its convection properties and potentially leading to enhanced material loss. While mass-loss rates in black widows and redbacks depend on multiple factors, including the intensity of irradiation and orbital evolution, models indicate that redback companions, being more massive, may sustain a stronger magnetic field, which provides greater resistance to ablation-driven mass loss \citep{Conrad_Burton_2023}.

The magnetic fields of redbacks exhibit larger fluctuations, with shifts of up to 150 kG, compared to the more stable 20 kG variations in black widows. However, the magnetic activity in redbacks is observed to persist for shorter timescales than in black widows (see Table \ref{tab:Table1}). This behaviour fits within the irradiation-driven evolution scenario, where the pulsar’s relativistic wind and high-energy radiation interact with the companion’s magnetosphere \citep{Conrad_Burton_2023}. In redbacks, the stronger magnetic field can counteract the pulsar wind for longer durations, delaying the complete stripping of the companion. In contrast, the weaker fields of black widow companions result in a more effective removal of material, leading to more rapid and extreme ablation \citep{Podsiadlowski1991,Chen2013,Conrad_Burton_2023}. Interestingly, while ablation is commonly associated with more severe mass loss in black widows, it is expected that redbacks' companions can intercept a larger fraction  of the pulsar’s spin-down luminosity, possibly as a result of a different geometric configuration \citep{Chen2013,Conrad_Burton_2023}. This effect likely stems from the more substantial convective envelope and deeper energy deposition in redback companions, which alters their thermal and magnetic structures.

From a modeling perspective, a convincing explanation of the connection between the directly observed surface and bulk magnetic field properties of these systems is still missing. The state of the art on the most relevant studies on this topic follows two distinct approaches, mainly based on three-dimensional MHD simulations \citep{simul_2008,Rakesh,macdonald} and one-dimensional stellar evolution analyses \citep{Gregory}. We can see that in the aforementioned works there is an ongoing disagreement about the order of magnitude of the field intensity in fully convective stars: some authors require $\sim$MG fields, while others insist on possible upper limits of $\sim$10 kG. Given this lack of consensus, our 20-100 kG subsurface field variation (see Fig. \ref{fig:magnetic}) at a depth $\sim0.1 R_c$ \citep{Applegate1992} seems to be reasonable. Currently, there are no simulations specifically focused on this aspect and up to now there are only some indications that the subsurface magnetic field could be larger than the surface one by a factor of $\sim2$ \citep[see Fig. 4 in][for details]{Rakesh}. 

One of the novel aspects of our model with respect to the literature relies on the temporal dynamics of the $\Delta B(t)$ plot. The long-term evolution of the surface fields is sometimes observed, while the suggested drastic variation of the field intensity by a factor of $\sim$10 with a characteristic oscillatory shape seems to be not seen in any other observations or theoretical modeling of fully-convective stars outside spider systems. It is true that our profile is based on a naive formula of the magnetic field, but the conventional dynamo models interplay with the quadrupole variations in these stars, possibly generating new behaviours. 

Regarding luminosity, redback companions generally exhibit higher optical luminosities than black widows, as shown in Fig. \ref{fig:luminosity}. This trend arises from a combination of factors \citep{Roberts2012,Gentile2014,Roberts2018,Strader2019,Sullivan2024}: (i) the intrinsic luminosity of the star, which is modulated by pulsar heating; (ii) the pulsar wind interaction with the companion and the surrounding material, particularly in X-ray bands; (iii) the gamma-ray emission from pulsar magnetospheric processes such as curvature radiation and inverse Compton scattering; (iv) occasional mass transfer episodes, where infalling matter is energized by the pulsar’s magnetic field; and (v) non-thermal radiation from the intra-binary shock formed between the pulsar wind and the ablated material.

In our case, redbacks reach total luminosities of $\mathcal{L}_0 \sim 10^{35}$ erg/s, while black widows are typically a few orders of magnitude dimmer ($\mathcal{L}_0 \sim 10^{31}$ erg/s). The larger companion star and stronger irradiation in redbacks contribute to a higher optical luminosity. However, in the X-ray and gamma-ray bands, black widows can still exhibit comparable or even greater luminosities due to the more efficient formation of intra-binary shocks \citep{Conrad_Burton_2023}.
These findings suggest that spider systems provide a unique laboratory for studying the effects of extreme irradiation on stellar magnetism, convection, and mass-loss processes. Future observational constraints and detailed simulations will be crucial in further refining our understanding of the evolution of these exotic binary systems.

\subsection{Parameters' determination}
\label{sec:parameter-determination}
In the current study, it is of essential importance to understand how to determine the parameters associated with a spider system using our model, which is fully based on the Applegate mechanism. This process is critical for extracting detailed information about the gravitational source under investigation. Broadly speaking, two main approaches can be identified.

The first route relies on the fact that when a new spider source is discovered, we can generally fit the orbital period modulations, thus obtaining the parameters $T_0$, the \texttt{TEMPO} coefficients $\{f_i\}_{I=0}^n$, and $P_{\rm ma}$. Now, assuming that the NS has a canonical mass of $m_p=1.4M_\odot$ and that the system is observed nearly edge-on, the mass of the companion star, $m_c$, can then be estimated \citep[see e.g.,][for more details]{Ridolfi2016,Freire2017}. This step is crucial to determine the spider system nature, namely whether it is a redback or a black widow. Subsequently, we can estimate the companion star's radius, $R_c$, by adopting the same argument detailed in Appendix \ref{appendix}. Using Kepler’s third law, the initial orbital separation, $a$, can then be calculated. By applying these criteria, we obtain a comprehensive measurement of all the independent parameters.

An alternative strategy shares with the above procedure the determination of the parameters $T_0$, $\{f_i\}_{i=0}^n$, and $P_{\rm ma}$ from the observations, while leaving the remaining four parameters $m_p,m_c,R_c,a$ to be inferred. They could be potentially determined if we have experimental data pertaining to a physical observable, such as for example the luminosity profile \eqref{eq:luimnosity}, which encapsulates a combination of all these unknowns. By fitting such data and extracting the best-fit parameters, with the annex constraints on the variation ranges and the Kepler's third law validity, a consistent set of values can be extracted. Depending on the available data, one or the other strategy could be used. Obviously, the available data varies from system to system.

\section{Conclusions}
\label{sec:end}
This article deals with spider binary systems, formed by a pulsar and a low-mass companion star, classified in redbacks ($m_c\sim0.1-0.4M_\odot$) and black widows ($m_c\ll 0.1M_\odot$). One of their distinctive features on which we concentrate is the long-term unpredictable variations in the orbital period and its first derivative \citep{Roberts2014}. Among the different contributions, which can account for orbital period modulations, it has been clearly shown that the most reasonable explanation is due to the Applegate mechanism \citep{Applegate1987,Applegate1992,Applegate1992pro}. This description accounts for orbital timing variations via the companion star's quadrupole moment changes induced by the magnetic dynamo action, which in turn is communicated to the orbital motion through quadrupole-gravity coupling. 

This paper constitutes the first attempt to make the Applegate mechanism dynamical. To the best of our knowledge, we  provide for the first time the spider observables' evolution. Combining information derived by pulsar timing observations and assuming the Applegate mechanism as true, we are able to track the evolution in time (within a determined timeframe) of some physical quantities (cf. Eq. \eqref{eq:output}).

\citet{Voisin2020a,Voisin2020b} considerably improved the treatment of the quadrupole deformations while also adding relativistic effects, but they only provided the dynamical evolution of the quadrupole moment. Another important difference between \citet{Voisin2020a,Voisin2020b} and our approach relies on the final goals. \citet{Voisin2020a,Voisin2020b} proposed a detailed model for describing the motion of spider binary systems to accurately estimate the observed $\Delta P$ variations with the ultimate objective of improving the timing solution pertaining to these gravitational sources. 

On the other side, our work employs a reverse approach: rather than finding a physical justification for the $\Delta P$ variations, we take them from the long-term observations and, by relying on the validity of the Applegate mechanism, we reconstruct the dynamics of the related physical variables. Therefore, we follow an observational and deductive analysis instead of a theoretical and inductive one. Even though our model is very simple, we emphasize that it could also be applied to more refined frameworks.
 
To achieve our goal, we have first derived the equations of motion \eqref{eq:EoM1} -- \eqref{eq:EoM2} pertaining to the dynamics of spider binaries, based on the Applegate works. Then, we have considered the function $\Delta P(t)$, reconstructed by fitting the observational data on the orbital period modulations. However, the resolution of this problem can be accomplished most likely only numerically and this can be excessively time consuming. To this end, we have developed a mathematical procedure (see Sec. \ref{sec:approx_meth}), based on approximating the quasi-circular orbit with a circular one, even though it presents mild oscillations on small timescales. This strategy has allowed us to obtain $r_{\rm S}(t)$ without solving Eq. \eqref{eq:EoM1}. Substituting this function and $\Delta P(t)$ into Eq. \eqref{eq:EoM2} we have that the ensuing differential equation is still difficult to solve analytically. 

Therefore, to avoid any kind of numerical integration, we have approximated $r_{\rm S}(t)$ and $\Delta P(t)$ with high-accurate polynomials. This has allowed us to have an analytical expression of $Q(t)$, whose functional formula has been reported in Eq. \eqref{eq:func_form_Q}. This result is in good agreement with the corresponding function computed by numerically integrating together Eqs. \eqref{eq:EoM1} -- \eqref{eq:EoM2} without making any simplifying hypothesis. Through these formulae, it has been possible to easily obtain the evolution of the related physical observables, which are (see Sec. \ref{sec:observables}): orbits, quadrupole moment, magnetic field, and luminosity. 

In Fig. \ref{fig:Fig3}, we have displayed the orbital period modulations of the redback 47 Tuc W and black widow 47 Tuc O, whereas in Figs. \ref{fig:radius}, \ref{fig:quadrupole}, \ref{fig:magnetic}, and \ref{fig:luminosity} we have shown the evolution of the related physical observables via our model. We have discussed analogies and differences among the two sources (see Sec. \ref{sec:comparison}), being representatives of the redbacks and black widows. We have contextualized these results in a more general physical picture.

The \emph{advantages} of our approach are: (1) making use of simple formulae; (2) having the evolution of the above described physical variables, which allow to be better modeled through future upgraded descriptions; (3) having insight into the subsurface magnetic activity inside the companion star, which is still not a clear topic; (4) employing our analytical formulae to tightly constraint the model parameters by using not only the orbital period modulations, but also the profile of other observables \citep[e.g.,][use the luminosity profile as further informaion]{Zhao2023}; (5) the vast application of our developments to more updated or different descriptions of spider binary systems. 

However, our treatment possesses also some evident \emph{limits}: (1) the model is very simple and must be improved under different naive aspects; (2) the magnetic field necessitates to be modeled through more realistic formulae, based on MHD simulations \citep{simul_2006,simul_2008}, to clarify the link between surface and subsurface magnetic fields, still not fully treated \citep{Morin2012}; (3) it is not adequate to reproduce dissipative phenomena \citep{Lanza2006}, as well as new effects as the general relativistic corrections and the three-dimensional motion of the two-bodies \citep{Voisin2020a,Voisin2020b}; (4) the model prediction power is limited only within the observational period $[t_0,t_1]$. 

The future perspectives can branch out into several routes. First, the methodology and results of this article could be applied to catalog all the available spider systems and to perform more accurate analyses, in order to extract relevant information on stellar evolution, and to classifyi the pulsar population in a more methodical fashion. Another possible development is to derive a more handy equation from actual MHD simulations to better model the magnetic activity inside the companion star.

\section*{Acknowledgments}
The authors greatly thank Oleg Kochukhov for valuable comments on our results. V.D.F. is grateful to Gruppo Nazionale di Fisica Matematica of Istituto Nazionale di Alta Matematica (INDAM) for support. V.D.F. acknowledges the support of INFN {\it sez. di Napoli}, {\it iniziativa specifica} TEONGRAV. V.D.F. is grateful to both the SRT -- Sardinia Radio Telescope and the Max Planck Institute für Radioastronomie in Bonn for the hospitality. A.Ca. is grateful to Scuola Superiore Meridionale for hospitality. A.R. is supported by the Italian National Institute for Astrophysics (INAF) through an `IAF - Astrophysics Fellowship in Italy' fellowship (Codice Unico di Progetto: C59J21034720001; Project `MINERS'). AR also acknowledges continuing valuable support from the Max-Planck Society.

\appendix
\section{Polynomial approximations}
\label{appendix}
We provide the polynomial approximations of $\Delta P(t),r_{\rm S}(t),Q(t)$ with the related MRE for the spider systems 47~Tuc~W (see Eq. \eqref{sec:47_TUC_W_pol}) and 47~Tuc~O (see Eq. \eqref{sec:47_TUC_O_pol}). The input parameters of these two physical systems are reported in Table \ref{tab:Table1}. 

We assume that the pulsar mass is the standard value $m_p=1.4 M_\odot$. The companion star's mass, $m_c$ is taken from works cited in the caption of Table \ref{tab:Table1}, assuming that the binary system is seen by the observer almost edge on, namely $\sin i =1$ with $i$ inclination of the observer with respect to the $z$-axis. Instead, the initial binary separation $a$ is calculated via the Kepler's third law (cf. Eq. \eqref{eq:unbert-period}), employing $f_0=1/P_0,m_p,m_c$.

Regarding the companion star's radius (at rest), $R_c$, there is no guidance on its calculation in the referenced papers on the two sources. To estimate it, we follow this strategy. In black widows, the companion can reach extremely low masses, as seen in 47 Tuc O, suggesting it may be a brown dwarf, whose radius typically falls within the range $0.064-0.113 R_\odot$ \citep{Sorahana2013}. For our purposes, we adopt an average radius of $R_c=0.08 R_\odot$. In contrast, for redbacks, the companion is a main-sequence star, allowing us to estimate its radius using the following formula \citep[see Table \ref{tab:Table1} and ][]{Demory2009}:
\begin{equation}
R_c=\left(\frac{m_c}{M_\odot}\right)^{0.8}R_\odot =0.20 R_\odot.   
\end{equation}
The radii we have selected are in agreement with the Roche lobe size $R_L$ of the two sources \citep{Frank2002}, as for 47 TUC W we have $R_c/R_L=0.78$, while for 47 TUC O it is $R_c/R_L=0.63$.

\begin{table*}[th!]
\centering
\caption{We show the input parameters \eqref{eq:input} pertaining to 47 Tuc W and 47 Tuc O, as well as the quantities $Q_0$ (cf. Eq. \eqref{eq:ICN}), $B_0$ (cf. Eq. \eqref{eq:magfield_App}), and $\mathscr{L}_0$ (cf. Eq. \eqref{eq:luminsity_fix}). We express all masses and distances in terms of the solar mass $M_\odot=2\times 10^{33}{\rm g}$ and the solar radius $R_\odot=6.96\times10^{10}{\rm cm}$, respectively. The input parameters and \texttt{TEMPO} coefficients $t_0,t_1,T_0,\{f_i\}_{i=0}^n$ are both taken from \citet{Ridolfi2016,Freire2017}. For more details on how $a$ and $R_c$ have been calculated/chosen, we refer to the beginning of Appendix \ref{appendix}.}
\begin{tabular}{|c|c|c|c|}
\hline
Parameter & Unit & 47 Tuc W & 47 Tuc O\\
\hline
$m_p$  & $M_\odot$ & 1.40 & 1.40\\
$m_c$  & $M_\odot$ & 0.13 & 0.022\\
$R_c$  & $R_\odot$ & 0.20 & 0.08\\
$a$  & $R_\odot$ & 1.26 & 1.11\\
$P_{\rm ma}$ & yr & 2.70 & 5.00\\
$Q_0$ & $M_\odot\ R_\odot^2$ & $3.61\times10^{-5}$ &$6.25\times10^{-6}$\\
$B_0$ & kG & 48.79 & 0.92 \\
$\mathscr{L}_0$ & $L_\odot$ & 34.78 & $4.81\times10^{-3}$\\
\hline
\multicolumn{4}{|c|}{}\\
\multicolumn{4}{|c|}{\texttt{TEMPO} coefficients}\\
\multicolumn{4}{|c|}{}\\
\hline
$t_0$ & MJD & 51214.216 &47717.894\\
$t_1$ & MJD & 54934.047 &56388.106\\
$T_0$  & MJD & 51585.3327 & 51600.1084 \\
$f_0$  &  s${}^{-1}$ & $8.71\times10^{-5}$ &$9.59\times10^{-5}$\\
$f_1$  & s${}^{-2}$ & $-1.27\times10^{-18}$ &$-2.09\times10^{-21}$ \\
$f_2$  & s${}^{-3}$ & $4.06\times10^{-26}$ &$-1.89\times10^{-28}$\\
$f_3$  & s${}^{-4}$ & $6.30\times10^{-33}$ &$3.94\times10^{-35}$\\
$f_4$  & s${}^{-5}$ & $-9.18\times10^{-40}$ &$-1.49\times10^{-43}$\\
$f_5$  & s${}^{-6}$ & $6.27\times10^{-47}$ &$-5.49\times10^{-50}$\\
$f_6$  & s${}^{-7}$ & $-2.68\times10^{-54}$ &$5.76\times10^{-58}$\\
$f_7$  & s${}^{-8}$ & $7.41\times10^{-62}$ &$5.478\times10^{-65}$\\
$f_8$  & s${}^{-9}$ & $-1.22\times10^{-69}$ &$-8.91\times10^{-73}$\\
$f_9$  & s${}^{-10}$& $9.30\times10^{-78}$ &$-3.90\times10^{-80}$\\
$f_{10}$ & s${}^{-11}$ & -- &$8.57\times10^{-88}$\\
$f_{11}$ & s${}^{-12}$ &-- &$1.52\times10^{-95}$\\
$f_{12}$ & s${}^{-13}$ &-- &$-4.21\times10^{-103}$\\
\hline
\end{tabular}
\label{tab:Table1}
\end{table*}

\subsection{Redback: 47 Tuc W}
\label{sec:47_TUC_W_pol}
We approximate $\Delta P(t)$ with the polynomial
\begin{align}\label{sec:P1}
&\mathscr{P}_1(t)=-1.24\times10^5 t^9+5.67\times10^5 t^8-1.10\times 10^6 t^7\notag\\
&+1.16\times 10^6 t^6-7.31\times10^5 t^5+2.80\times10^5 t^4\notag\\
&-6.30\times10^4 t^3+7.62\times10^3 t^2-3.97\times10^2 t+4.74,    
\end{align}
where $t\in[0,1]$, which can be cast in $[t_0,t_1]$ via Eq. \eqref{eq:time-transformation}, and $\mathscr{P}_1(t)$ has the dimension of ms. The related MRE is $\sim10^{-5}\%$, which is extremely accurate, since we have used a ninth-order polynomial (see discussion of Sec. \ref{sec:succ_stra}, for details). 

We use the approximation \eqref{sec:P1} for calculating the radius $r_{\rm S}(t)$ (cf. Eqs. \eqref{eq:definitions} and \eqref{delta_r}), committing a MRE of $\sim10^{-13}\%$, which is still very accurate. For the quadrupole moment, we employ Eq. \eqref{eq:Q} and this polynomial approximation for $r_{\rm S}(t)$
\begin{align}
&\mathscr{R}_1(t)=-6.31\times 10^8 t^9+2.89\times 10^9 t^8-5.58\times 10^9 t^7\notag\\
&+5.91\times 10^9 t^6-3.73\times 10^9 t^5+1.43\times 10^9 t^4-3.21\times 10^8 t^3\notag\\
&+3.89\times 10^7 t^2-2.02\times 10^6 t+8.79\times 10^{10},    
\end{align}
where $t\in[0,1]$ and $\mathscr{R}_1(t)$ is expressed in cm. The MRE for $r_{\rm S}(t)$ is of $\sim10^{-13}\%$. Using this approach, the MRE on $Q(t)$ is $\sim\times10^{-3}\%$, which is still in good agreement. We conclude that the quadrupole moment is approximated with respect to the original solution with a MRE of $0.004\%$. 

\subsection{Black widow: 47 Tuc O}
\label{sec:47_TUC_O_pol}
We approximate $\Delta P(t)$ with the polynomial (in ms unit)
\begin{align}\label{sec:P2}
&\mathscr{P}_2(t)=2.98\times 10^6 t^{12}-1.78\times 10^7 t^{11}+4.66\times 10^7 t^{10}\notag\\
&-7.07\times 10^7 t^9+6.88\times 10^7 t^8-4.48\times 10^7 t^7\notag\\
&+1.98\times 10^7 t^6-5.89\times 10^6 t^5+1.15\times 10^6 t^4\notag\\
&-1.40\times10^5 t^3+9.72\times10^3 t^2-3.42\times10^2 t+4.59,  
\end{align}
where $t\in[0,1]$, whose related MRE is $\sim\times10^{-3}\%$, being very accurate, as $\mathscr{P}_2(t)$ is of twelfth order (see Sec. \ref{sec:succ_stra}, for details). 

The approximation \eqref{sec:P2} is exploited for computing the radius $r_{\rm S}(t)$ with a MRE of $\sim10^{-14}\%$. The estimation of the quadrupole moment \eqref{eq:Q} is performed by approximating the radius $r_{\rm S}(t)$ via the following polynomial (in cm unit and $t\in[0,1]$)
\begin{align}
&\mathscr{R}_2(t)=1.48\times 10^{10} t^{12}-8.79\times 10^{10} t^{11}+2.30\times 10^{11} t^{10}\notag\\
&-3.50\times 10^{11} t^9+3.41\times 10^{11} t^8-2.22\times 10^{11} t^7\notag\\
&+9.81\times 10^{10} t^6-2.92\times 10^{10} t^5+5.69\times 10^9 t^4-6.91\times 10^8 t^3\notag\\
&+4.81\times 10^7 t^2-1.69\times 10^6 t+7.74\times 10^{10}.    
\end{align}
The associated MRE is of $\sim10^{-12}\%$. Therefore, the MRE on $Q(t)$ is $\sim10^{-4}\%$, being in perfect agreement with the original formula. Finally, the quadrupole moment is approximated with respect to the original solution with a MRE of $0.03\%$. 

\bibliographystyle{aa}
\bibliography{references}

\end{document}